\documentclass[12pt]{iopart}

\usepackage{graphicx}
\eqnobysec

\begin{document}

\begin{flushright}
KUNS-2383
\end{flushright}

\title[Statistical Anisotropy from Anisotropic Inflation]{Statistical Anisotropy from Anisotropic Inflation  }

\author{Jiro Soda}

\address{Department of Physics,  Kyoto University, Kyoto, 606-8502, Japan}
\ead{jiro@tap.scphys.kyoto-u.ac.jp}
\begin{abstract}
We review an inflationary scenario with the anisotropic expansion rate. 
An anisotropic inflationary universe can be realized by a vector field coupled with an inflaton, which
can be regarded as a counter example to the cosmic no-hair conjecture. 
We  show generality of anisotropic inflation and derive a universal property. 
We formulate cosmological perturbation theory in anisotropic inflation.
Using the formalism, we show anisotropic inflation gives rise to the statistical anisotropy
in primordial fluctuations. We also explain a method to test anisotropic inflation using the cosmic
microwave background radiation (CMB). 
\end{abstract}

\maketitle
\section{Introduction}

It is well known that inflation elegantly solves the horizon and flatness problems.
Moreover, inflation accounts for the origin of the large scale structure of the universe.
The point in an inflationary scenario is that the exponential expansion of the universe
erases any classical memory, which is often referred to as the cosmic no-hair conjecture. 
Because of this feature,  quantum fluctuations are responsible for the origin of the large scale structure
of the universe. Remarkably, the nature of the primordial fluctuations is understood by symmetry
in inflation. 
\begin{itemize}
\item {\bf homogeneity}\\
First of all, we have to assume the initial homogeneity. 
Indeed, inflation does not commence with strong inhomogeneous initial conditions
( There is a possibility that this symmetry breaks down~\cite{ArmendarizPicon:2007nr,Libanov:2010nk,arXiv:1111.2721}.).

\item {\bf shift symmetry}\\
In order to have slow-roll inflation, we need a sufficiently flat potential.
Hence, we have a shift symmetry in field space, $\phi (x) \rightarrow \phi (x) + \bar{\phi} $.
Here, $\bar{\phi}$ is a constant. 

\item {\bf temporal de Sitter symmetry}\\
The metric for de Sitter spacetime reads 
\begin{eqnarray}
 ds^2 = -dt^2 + e^{2 Ht} ( dx^2 + dy^2 + dz^2) \ ,
\end{eqnarray}
where the Hubble parameter $H$ is constant. It is easy to find isometry
$$
t \rightarrow t+ \bar{t} \ ,\quad  x^i \rightarrow e^{-H \bar{t}} x^i \ , \quad x^i =\{x,y,z\} \ ,
$$
where $\bar{t}$ is a constant.

\item {\bf spatial de Sitter symmetry}\\
 Once inflation occurs, the cosmic no-hair conjecture tells us that
the universe will be isotropized in a few Hubble expansion. 
\end{itemize}
Here, we assumed single field inflation with a standard kinetic term. 

The above symmetry determines the nature of primordial fluctuations.
In general, we need $n$-point correlation functions to characterize the statistical nature of
primordial fluctuations. However, the {\sl shift symmetry} in field space 
implies suppression of nonlinearity and hence the Gaussian statistics of fluctuations.
Thus, we need only 2-point functions. As an example, we take curvature perturbations $\zeta$.
In a Fourier space, we have the power spectrum
$$< \zeta ({ \bf k}_1 )  \ ,  \zeta ({ \bf k}_2 ) > = P ({ \bf k}_1 \ , { \bf k}_2) \ , $$
where ${\bf k}$ is the wavenumber vector. 
Moreover, the {\sl homogeneity} constrains the power spectrum to be 
$$< \zeta ({ \bf k}_1 )  \ ,  \zeta ({ \bf k}_2 ) > = \delta ({ \bf k}_1 + { \bf k}_2) P ({\bf k}_1) . $$
Here, the delta function stems from the ``momentum" conservation. 
The {\sl spatial de Sitter symmetry} further constrains the power spectrum as   
$$< \zeta ({ \bf k}_1 )  \ ,  \zeta ({ \bf k}_2 ) >
 = \delta ({ \bf k}_1 + { \bf k}_2) P (k_1 = |{\bf k}_1 | ) . $$
Namely, the direction dependence is forbidden by the rotational symmetry. 
Finaly, the {\sl temporal de Sitter
symmetry} yields a scale invariant power spectrum
$$ P({\bf k}) = {\rm const.} $$ 
because of the scale invariance of spatial coordinates.  
These predictions are robust and universal in inflationary scenarios.
In fact, the above predictions have been confirmed by CMB observations~\cite{Komatsu:2010fb}.

However, precision cosmology forces us to look at fine structure of the primordial fluctuations.
In fact, since the universe is not exactly de Sitter, there exists violation of the temporal
de Sitter symmetry, which leads to a slight tilt of the power spectrum. As the deviation from 
de Sitter expansion can be characterized by the slow roll parameter, the tilt should be 
of the order of the slow roll parameter. Similarly, we have violation of the shift symmetry,
which leads to non-Gaussianity characterized by the slow roll parameter~\cite{Maldacena:2002vr}. 
Along this line of thought,
it is natural to expect violation of the spatial de Sitter symmetry, which would lead to the statistical
anisotropy. 

From the observational point of view, a lot of anomalies indicating the statistical anisotropy are reported
although its statistical significance is uncertain ( see \cite{Bennett:2010jb} and references therein).
Motivated by those observations, there are many theoretical proposals
to realize the statistical anisotropy~\cite{Yokoyama:2008xw,Karciauskas:2008bc,Dimopoulos:2009am,
Dimopoulos:2009vu,ValenzuelaToledo:2009af,ValenzuelaToledo:2009nq,arXiv:1107.3186,arXiv:1108.4424,Bartolo:2009pa,
Bartolo:2009kg,Dimastrogiovanni:2010sm,arXiv:1111.1919,arXiv:1111.1929,arXiv:1104.3629,arXiv:1107.2779,BeltranAlmeida:2011db}. 
In  these works, however, a consistent theoretical framework including the backreaction of vector fields seems to be obscure. 
As another line of research, there exist  challenges to the cosmic no-hair conjecture~\cite{Ford:1989me,Kaloper:1991rw,Barrow:2005qv,
Barrow:2009gx,Campanelli:2009tk,arXiv:1103.6175,Golovnev:2008cf,arXiv:1109.4838,Kanno:2008gn}.  
If we can evade the cosmic no-hair conjecture, we would have the statistical anisotropy. 
Unfortunately, it turns out that these models suffer from either the instability,
or a fine tuning problem, or a naturalness problem~\cite{Himmetoglu:2008zp}.
Recently, however, stable anisotropic inflation models are found in the context of supergravity,
which gives rise to a counter example to the cosmic no-hair conjecture~\cite{Watanabe:2009ct}.  

In this review, we explain how anisotropic inflation can be realized in supergravity and
show it implies violation of spatial de Sitter symmetry and hence leads to the statistical
anisotropy in primordial fluctuations.
 Recall that the bosonic sector of the supergravity action is given by
\begin{eqnarray}
  \hspace{-2cm}
S &=& \int d^4 x \sqrt{-g} \left[ \frac{1}{2\kappa^2} R - G_{\bar{i}j} \partial^\mu \bar{\phi}^{\bar{i}} \partial_\mu \phi^j 
             -e^{\kappa^2 K} \left( G^{\bar{i}j} \bar{D}_i \bar{W} D_j W - 3\kappa^2 \bar{W} W \right) \right.\nonumber\\
\hspace{-2cm}
   && \left. \hspace{5cm} - \frac{1}{4} f^2_{ab}(\phi ) F^{a\mu\nu} F^b_{\mu\nu} +\cdots \hspace{1cm} \right] \ ,
\end{eqnarray}
where $G_{\bar{i}j}=\partial K /\partial \phi^{\bar{i}} \partial \phi^j$, $D_i W = \partial W /\partial \phi^i +\kappa^2 (\partial K/\partial \phi^i ) W$, $K(\phi ,\bar{\phi})$ and $W(\phi)$ are the Kaler potential and the super potential, respectively. 
There is also a kinetic term for gauge fields with gauge kinetic functions $f_{ab}$.
The cosmological role of $K$ and $W$ are well discussed so far. However, the gauge kinetic function
has been overlooked in cosmology, at least in the context of inflationary scenarios. 
In this review, we clarify the role of the gauge kinetic functions in inflation.

Let us summarize main results here. First of all, we find anisotropic inflation is an attractor
in supergravity with a wide rage of gauge kinetic functions. The metric during inflation approximately reads
\begin{eqnarray}
   ds^2 = -dt^2 + e^{2Ht} \left[ e^{-4\Sigma t} dx^2 + e^{2\Sigma t} \left( dy^2 + dz^2 \right) \right] \ ,
\end{eqnarray}
where $H$ and $\Sigma$ describe the average expansion rate and the anisotropic expansion rate, respectively.  
The degree of the anisotropy $\Sigma / H $ is universally given by the following formula
\begin{eqnarray}
  \frac{\Sigma}{H} = \frac{1}{3} I \epsilon_H \ , \quad \epsilon_H = - \frac{\dot{H}}{H^2} \ ,
\label{main-formula}
\end{eqnarray}
where $I$ is the model parameter taking values $0\leq I \leq 1$. The point is that the degree of anisotropy
at most of the order of the slow-roll parameter $\epsilon_H $. 
In this scenario, we have the statistical anisotropy of the form~\cite{Ackerman:2007nb}
\begin{eqnarray}
P({\bf k}) = P(k) \left[ 1+ g_* \sin^2 \theta \right] \ ,
\end{eqnarray}
where $P(k)$ is the isotropic part of the power spectrum $P({\bf k})$ and
 $\theta$ is the angle between the preferred direction and the wavenumber vector of fluctuations.
The amplitude of anisotropy $g_*$ can be calculated using the standard perturbation theory.
The anisotropy in curvature perturbations is given by
\begin{eqnarray}
g_s = 24 I N^2 (k) 
\end{eqnarray}
 and that in gravitational waves reads
\begin{eqnarray}
g_t = 6 I \epsilon_H N^2 (k) \ .
\end{eqnarray}
Here, $I$ is a model parameter appeared in the anisotropy formula (\ref{main-formula})
and $N(k)$ is the $e$-folding number from the horizon exit of fluctuations to the end of inflation.
Remarkably, there exists a difference between the anisotropy in curvature and tensor perturbations.
There is also the cross-correlation between the curvature perturbations $\zeta$ and gravitational waves $h$
given by 
\begin{eqnarray} 
r_c = \frac{<\zeta h>}{<\zeta \zeta >} = -24 \sqrt{2} I \epsilon_H N^2 (k) . 
\end{eqnarray}
We find the consistency relations between these observables
\begin{eqnarray}
   4 g_t = \epsilon_H \  g_s   \ , \quad r_c = -\sqrt{2} \epsilon_{H} g_s  \ .
\end{eqnarray}
This allows us to give a model independent test of anisotropic inflation.
Of course, we can use each observables to constrain gauge kinetic functions.
Indeed, we give the first cosmological constraint on the gauge kinetic function:
\begin{eqnarray}
 I < \frac{0.3}{24 N^2 (k)} \ ,
\end{eqnarray}
where we used the result in \cite{Pullen:2007tu}.
Note that $I$ is determined once the gauge kinetic functions are given.

The organization of the paper is as follows. In section II, we show
power-law inflation is not necessarily an attractor in the presence of gauge kinetic function.
Instead, we show anisotropic inflation could become an attractor for a wide range of parameters in models.
In section III, we argue the generality of the model and derive a universal relation.
In section IV, we develop cosmological perturbation theory in anisotropic inflation
and calculate the statistical anisotropy in primordial fluctuations.
In section V, we discuss observational test of anisotropic inflation using the CMB.
The final section is devoted to summary and future prospects.

\section{Primordial Magnetic Fields, Backreaction, and Anisotropic Inflation}
\label{sec:2}

In this section, we first recall a standard mechanism for generating primordial
magnetic fields during inflation. There, a vector field coupled to an inflaton is introduced.
Here, we consider a simple model with exponential
potential and gauge kinetic functions. Then we obtain exact power-law inflation.
It is possible to generate primordial magnetic fields in this set up.
For many cases, the backreaction of the vector field cannnot be negligible. 
Indeed, it turns out that the backreaction yields a new type of cosmological solutions, namely, 
exact anisotropic power-law inflation.

\subsection{Power-law Inflation and Primordial Magnetic Fields}

 Let us start with the action
\begin{eqnarray}
S&=&\int d^4x\sqrt{-g}\left[~\frac{1}{2\kappa^2}R
-\frac{1}{2}\left(\partial_\mu\phi\right)\left(\partial^{\mu}\phi\right)  -V(\phi) 
~\right] \ ,
\label{eq:action1}
\end{eqnarray}
where $\kappa ^2$ is the reduced gravitational constant, $g$ is the determinant of the metric, $R$ is the Ricci scalar, 
$V(\phi)$ is a potential for an inflaton $\phi$.
In order to find exact solutions, we take the exponential potential
\begin{eqnarray}
V (\phi) = V_0 e^{\lambda \kappa \phi } \ ,
\label{V}
\end{eqnarray}
where $V_0$ and $\lambda$ are  parameters.  It is natural to take
the isotropic metric
\begin{equation}
ds^2 = - dt^2 + e^{2\alpha(t)} \left[ dx^2 + dy^2 + dz^2 \right] \ ,
\end{equation} 
where $e^\alpha$ is the scale factor. 
Let us seek isotropic power-law solutions by putting the ansatz
\begin{eqnarray}
  \alpha = \zeta \log t \ , \qquad 
  \kappa \phi = \xi \log t + \phi_0 \ .
\end{eqnarray}
Then, we obtain the solutions
\begin{eqnarray}
  \zeta = \frac{2}{\lambda^2} \ , \hspace{1cm} 
  \xi = - \frac{2}{\lambda} \ , \hspace{1cm} 
  \kappa^2 V_0 e^{\lambda \kappa \phi_0} = \frac{2(6-\lambda^2)}{\lambda^4} \ .
  \label{iso-power}
\end{eqnarray}
In this case, we have the spacetime
\begin{eqnarray}
 ds^2 = -dt^2 + t^{4/\lambda^2} \left( dx^2 +dy^2 + dz^2 \right) \ .
 \label{isotropic}
\end{eqnarray}
Thus, for $\lambda \ll 1$, we have power-law inflation. 

Now, we consider primordial magnetic fields in this inflationary background.
We introduce a vector field $A_{\mu}$ whose kinetic term is 
coupled to the inflaton field $\phi$
\begin{eqnarray}
S&=&\int d^4x\sqrt{-g}\left[~-\frac{1}{4} f(\phi )^2 F_{\mu\nu}F^{\mu\nu}  
~\right] \ ,
\label{eq:action2}
\end{eqnarray}
where $F_{\mu\nu}$ is the field strength of the vector field defined by $F_{\mu\nu}=\partial_{\mu}A_{\nu}-\partial_{\nu}A_{\mu}$, and $f(\phi)$ is a coupling function of the vector field.
We emphasize that this kind of model is quite natural in the context of
the supergravity~\cite{Martin:2007ue}. 
Many years ago, Ratra considered the exponential gauge kinetic function~\cite{Ratra:1991bn}
\begin{eqnarray}
f(\phi) = f_0 e^{\rho \kappa \phi } 
\label{f}
\end{eqnarray}
and concluded that the primordial magnetic fields can be generated due to
the gauge kinetic function. The result implies there exists a
contribution of the vector field to the energy density in the universe. 
In the subsequent two subsections, we would like to show the backreaction of the vector field 
leads to the anisotropic inflationary power-law solutions and prove that
anisotropic inflation is actually an attractor~\cite{Kanno:2010nr}.
Thus, the models producing primordial magnetic fields naturally lead to anisotropic inflation~\cite{Kanno:2009ei,Dimopoulos:2008em}.
Here, we should emphasize that the existence of anisotropic inflation attractor shows that the backreaction from the vector field 
does not necessarily destroy inflation~\cite{Kanno:2009ei}, as is often assumed in literature~\cite{Demozzi:2009fu}.

\subsection{Backreaction and Anisotropic Power-law Inflation}

Now, we take into account the backreaction of the vector field.
We can expect interesting effects due to the backreaction because the coupling of the vector field 
to the inflaton produces an effective potential for the inflaton~\cite{hep-th/0604192,Soda:2006wr}. 
Indeed, we will see that there exist exact anisotropic  
solutions~\cite{Kanno:2010nr} on the contrary to the expectation from the cosmic no-hair theorem~
\cite{Wald:1983ky,Moss:1986ud,Kitada:1991ih,Kitada:1992uh}.

The action we should consider is given by
\begin{eqnarray}
S&=&\int d^4x\sqrt{-g}\left[~\frac{1}{2\kappa^2}R
-\frac{1}{2}\left(\partial_\mu\phi\right)\left(\partial^{\mu}\phi\right)
-V(\phi)-\frac{1}{4} f(\phi )^2 F_{\mu\nu}F^{\mu\nu}  
~\right] \ ,
\label{eq:action}
\end{eqnarray}
where $V(\phi)$ and $f(\phi)$ are given by (\ref{V}) and (\ref{f}), respectively.
Without loosing the generality, one can take  
$x$-axis in the direction of the vector field. Using the gauge invariance,
we can express the vector field as 
\begin{eqnarray}
 A_{\mu}dx^{\mu} = v(t) dx  \ .
\end{eqnarray} 
Thus, there exists the rotational symmetry in the $y$-$z$ plane. 
 Given this configuration, it is convenient to parameterize the metric as follows:
\begin{equation}
\hspace{-2cm}
ds^2 = -{\cal N}(t)^2dt^2 + e^{2\alpha(t)} \left[ e^{-4\sigma (t)}dx^2
+e^{2\sigma (t)}(e^{2\sqrt{3}\sigma_{-}(t)}dy^2
+e^{-2\sqrt{3}\sigma_{-}(t)}dz^2) \right] \ ,
\end{equation} 
where $e^\alpha$, $\sigma$ and $\sigma_{-}$ are an isotropic scale factor  
 and spatial shears, respectively.
 Here, the lapse function ${\cal N}$ is introduced to obtain the Hamiltonian constraint. 
With the above ansatz, the action becomes
\begin{equation}
\hspace{-2cm} S=\int d^4x \frac{1}{\cal N}e^{3\alpha} \left[ \frac{3}{\kappa ^2} (-\dot{\alpha}^2+\dot{\sigma}^2
+\dot{\sigma}_{-}^2)+\frac{1}{2}\dot{\phi}^{2}-{\cal N}^2 V(\phi)+\frac{1}{2}f(\phi )^2\dot{v}^2 e^{-2\alpha (t) +4\sigma(t) } \right],
\end{equation}
where an overdot denotes a derivative with respect to the physical time $t$.
First, its variation with respect to $\sigma_{-}$ yields
\begin{equation}
\ddot{\sigma}_{-}=-3\dot{\alpha}\dot{\sigma}_{-} \ .
\end{equation}
This gives $\dot{\sigma}_{-}\propto e^{-3\alpha}$, hence,
 the anisotropy in the $y$-$z$ plane rapidly decays as the universe expands. 
 Hereafter, for simplicity, we assume $\sigma_{-}=0$ and set the metric to be 
\begin{eqnarray}
 ds^2 = -dt^2 + e^{2\alpha(t)} \left[ e^{-4\sigma (t)}dx^2
 +e^{2\sigma (t)}(dy^2+dz^2) \right] \ .
\end{eqnarray}
Next, the equation of motion for $v$ is easily solved as
\begin{equation}
\dot{v} =f(\phi)^{-2}  e^{-\alpha -4\sigma}  p_A  \ ,
\label{eq:Ax}
\end{equation}
where $p_A$ is a constant of integration. Taking the variation of the action with respect to 
${\cal N}, \alpha, \sigma$ and $\phi$ and substituting the solution (\ref{eq:Ax}) into them, 
we obtain the following basic equations:
\begin{eqnarray}
\dot{\alpha}^2 &=& \dot{\sigma}^2+\frac{\kappa ^2}{3} \left[ \frac{1}{2}\dot{\phi}^2+V(\phi )+\frac{p_A^2}{2}f(\phi )^{-2}e^{-4\alpha -4\sigma} \right] \ , \label{eq:hamiltonian}\\
\ddot{\alpha} &=& -3\dot{\alpha}^2+\kappa ^2 V(\phi ) +\frac{\kappa ^2 p_A^2}{6}f(\phi )^{-2}e^{-4\alpha-4\sigma} \ , \label{eq:alpha}\\
\ddot{\sigma} &=&  -3\dot{\alpha}\dot{\sigma}+\frac{\kappa ^2 p_A^2}{3}f(\phi)^{-2}e^{-4\alpha -4\sigma} \ , \label{eq:sigma}\\
\ddot{\phi} &=& -3\dot{\alpha}\dot{\phi}-V_{\phi}+p_A^2f(\phi)^{-3} f_{\phi}e^{-4\alpha -4\sigma} \label{eq:inflaton} \ ,
\end{eqnarray}
where the subscript in $V_\phi$ denotes a derivative with respect to $\phi$.
 Let us check whether inflation occurs in this model. Using Eqs. (\ref{eq:hamiltonian}) and (\ref{eq:alpha}), the equation for acceleration of the cosmic expansion is given by
\begin{equation}
\frac{(e^{\alpha})^{\cdot \cdot}}{e^{\alpha}} = \ddot{\alpha}+\dot{\alpha}^2 = -2\dot{\sigma}^2-\frac{\kappa ^2}{3} \dot{\phi}^2 + \frac{\kappa ^2}{3} \left[ V - \frac{p_A^2}{2}f^{-2}e^{-4\alpha -4\sigma} \right] \ .
\end{equation}
We see that the potential energy of the inflaton needs to be dominant and the energy density of the vector field 
\begin{eqnarray}
\rho _v \equiv \frac{1}{2} p_A^2 f(\phi)^{-2}e^{-4\alpha -4\sigma}
\end{eqnarray} 
and the shear $\Sigma \equiv \dot{\sigma}$ should be subdominant for inflation to occur.

In order to find exact solutions, we take the power-law ansatz
\begin{eqnarray}
  \alpha = \zeta \log t \ , \hspace{1cm}
  \sigma = \eta \log t \ , \hspace{1cm}
  \kappa \phi = \xi \log t + \phi_0 \ .
\label{ask}
\end{eqnarray}
From the hamiltonian constraint equation (\ref{eq:hamiltonian}),
we get two relations
\begin{eqnarray}
  \lambda \xi = -2  \ , \hspace{1cm} \rho \xi +2 \zeta + 2\eta =1
  \label{A}
\end{eqnarray}
to have the same time dependence for each term. 
The latter relation is necessary 
only in the non-trivial vector case, $p_A \neq 0$.
Then, for the amplitudes to be balanced, we need 
\begin{eqnarray}
  -\zeta^2 +\eta^2 +\frac{1}{6} \xi^2 + \frac{1}{3} u +\frac{1}{6} w =0 \ ,
  \label{B}
\end{eqnarray}
where we have defined variables 
\begin{eqnarray}
   u = \kappa^2 V_0 e^{\lambda \phi_0} \  ,  \qquad
   w = \kappa^2 p_A^2 f_0^{-2} e^{-2\rho \phi_0}\,.
\end{eqnarray}
The equation for the scale factor (\ref{eq:alpha}) 
under Eq.~(\ref{A}) yields
\begin{eqnarray}
  -\zeta + 3\zeta^2 -u - \frac{1}{6} w =0 \ .
  \label{C}
\end{eqnarray}
Similarly, the equation for the anisotropy (\ref{eq:sigma}) gives 
\begin{eqnarray}
  -\eta + 3 \zeta\eta - \frac{1}{3} w =0 \ .
  \label{D}
\end{eqnarray}
Finally, from the equation for the scalar (\ref{eq:inflaton}),
we obtain
\begin{eqnarray}
  -\xi + 3\zeta \xi + \lambda u -\rho w = 0  \ .
  \label{E}
\end{eqnarray}
Using Eqs.~(\ref{A}),~(\ref{C}) and (\ref{D}), we can solve $u$ and $w$ as
\begin{eqnarray}
u &=& \frac{9}{2} \zeta^2 - \frac{9}{4} \zeta - \frac{3\rho}{2\lambda} \zeta
                 + \frac{1}{4} + \frac{\rho}{2\lambda}  \ , \label{u}
\\
w &=& -9 \zeta^2 + \frac{15}{2} \zeta + \frac{9\rho}{\lambda} \zeta 
                 -\frac{3}{2} -\frac{3\rho}{\lambda} \,.
                 \label{w}
\end{eqnarray}
Substituting these results into Eq.~(\ref{E}), we obtain
\begin{eqnarray}
  \left( 3\zeta-1 \right) \left[ 6 \lambda \left( \lambda + 2\rho \right)\zeta 
  - \left( \lambda^2 + 8\rho \lambda + 12 \rho^2 + 8 \right) \right] =0 \ .
\end{eqnarray}
In the case of $\zeta=1/3$, we have $u=w=0$. Hence, it is not our desired 
solution. Thus, we have to choose
\begin{eqnarray}
 \zeta = \frac{\lambda^2 + 8 \rho \lambda + 12 \rho^2 +8}{6\lambda (\lambda + 2\rho)} \ .
\label{expansion-rate}
\end{eqnarray}
Substituting this result into Eq.~(\ref{C}), we obtain
\begin{eqnarray}
 \eta = \frac{\lambda^2 + 2\rho \lambda -4 }{3\lambda (\lambda + 2\rho)}\,. 
\end{eqnarray}
This clearly shows the existence of the anisotropy in the expansion.
From Eq.~(\ref{A}), we have
\begin{eqnarray}
 \xi = - \frac{2}{\lambda}\,.
\end{eqnarray}
Finally, Eqs.~(\ref{u}) and (\ref{w}) reduce to
\begin{eqnarray}
 u = \frac{(\rho \lambda + 2\rho^2 +2)(-\lambda^2 + 4\rho \lambda +12 \rho^2 +8)}
      {2\lambda^2 (\lambda +2\rho )^2 }
\end{eqnarray}
and
\begin{eqnarray}
 w = \frac{(\lambda^2 + 2\rho \lambda -4)(-\lambda^2 + 4\rho \lambda +12 \rho^2 +8)}
      {2\lambda^2 (\lambda +2\rho )^2 } \ .
\end{eqnarray}
Note that Eq.~(\ref{B}) is automatically satisfied.
Thus, we have obtained anisotropic power-law solutions.

 Recalling the definition (\ref{ask}), we see $\zeta \gg 1$ is necessary for inflation. 
From the solution (\ref{expansion-rate}), it turns out that this requirement can be achieved
by assuming $\lambda \ll \rho$.
For these cases, $u$ is always positive.
Since $w$ should be also positive, we have the condition
\begin{eqnarray}
\lambda^2 + 2\rho \lambda > 4 \ .
\end{eqnarray}
Hence, $\rho $ must be much larger than one. 
Now, the spacetime reads
\begin{eqnarray}
  ds^2 = -dt^2 + t^{2\zeta-4\eta} dx^2 + t^{2\zeta +2\eta} 
\left( dy^2 + dz^2 \right) \ .
  \label{anisotropic}
\end{eqnarray}
The average expansion rate is determined by $\zeta$ and the average slow roll parameter
is given by
\begin{eqnarray}
\epsilon_H \equiv -\frac{\dot{H}}{H^2}
= \frac{6\lambda (\lambda + 2\rho)}{\lambda^2 + 8 \rho \lambda + 12 \rho^2 +8}
\ ,
\label{epsilon}
\end{eqnarray} 
where we have defined $H=\dot{\alpha}$. 
In the limit  $\lambda \ll 1$ and $\rho \gg 1$, 
this reduces to $\epsilon_H = \lambda /\rho$.  
Now, the anisotropy is characterized by
\begin{eqnarray}
  \frac{\Sigma}{H} \equiv \frac{\dot{\sigma}}{\dot{\alpha}}
  = \frac{2(\lambda^2 + 2\rho \lambda -4) }
  {\lambda^2 + 8 \rho \lambda + 12 \rho^2 +8} \ . 
\label{soverh}
\end{eqnarray}
From Eq.~(\ref{epsilon}) and (\ref{soverh}), we obtain a relation
\begin{eqnarray}
\frac{\Sigma}{H} = \frac{1}{3}I\epsilon_H \,,
\hspace{1cm}
I=\frac{\lambda^2 + 2\rho\lambda - 4}{\lambda^2 + 2\rho\lambda}\,.
\end{eqnarray}
It is possible to write $I$ as
\begin{eqnarray}
I=\frac{c-1}{c}\,,
\hspace{1cm}
c=\frac{\lambda^2 + 2\rho\lambda}{4} \ .
\end{eqnarray}
Then, it is apparent $I$ takes a value in the range $0 <I<1$.
We see the anisotropy is positive and proportional to the slow roll parameter $\epsilon_H$.

Although the anisotropy is always small, it persists during inflation.
Clearly these exact solutions give rise to counter examples to the cosmic no-hair 
conjecture. We should note that the cosmological constant is assumed in
the cosmic no-hair theorem presented by Wald~\cite{Wald:1983ky}. 
In the case of isolated vacuum energy, the inflaton can mimic the 
cosmological constant. 
However, in the presence of the non-trivial coupling between the inflaton
and the vector field, the cosmic no-hair theorem cannot
be applicable anymore.   

\subsection{Anisotropic Inflation as an Attractor}

In the previous subsections, we found both isotropic and anisotropic power-law solutions exist.
In this subsection, we will investigate the phase space structure.
Then, we will see which one is dynamically selected.

Let us use e-folding number as a time coordinate $d\alpha = \dot{\alpha} dt$.
It is convenient to define dimensionless variables
\begin{eqnarray}
  X = \frac{\dot{\sigma}}{\dot{\alpha}} \ , \hspace{1cm}
  Y = \kappa \frac{\dot{\phi}}{\dot{\alpha}} \ , \hspace{1cm}
  Z = \kappa f(\phi) e^{-\alpha +2\sigma} \frac{\dot{v}}{\dot{\alpha}} \ .
\end{eqnarray}
With these definitions, we can write the hamiltonian constraint equation
as
\begin{eqnarray}
  - \kappa^2 \frac{V}{\dot{\alpha}^2}
  = 3(X^2 -1) + \frac{1}{2} Y^2 + \frac{1}{2} Z^2 \ . 
  \label{hamconst}
\end{eqnarray}
Since we are considering a positive potential, we have the inequality
\begin{eqnarray}
3(X^2 -1) + \frac{1}{2} Y^2 + \frac{1}{2} Z^2 < 0 \ .
\label{constraint}
\end{eqnarray}

Using the hamiltonian constraint (\ref{hamconst}), we can eliminate $\phi$
from the equations of motion. 
Thus, the equations of motion can be reduced to the autonomous form:
\begin{eqnarray}
\frac{dX}{d\alpha} &=& \frac{1}{3} Z^2 (X+1) 
+ X\left\{ 3(X^2 -1) + \frac{1}{2} Y^2 \right\} 
\label{eq:X} \,,\\
\frac{dY}{d\alpha} &=& (Y+\lambda) \left\{ 3(X^2 -1) + \frac{1}{2} Y^2 \right\}
+ \frac{1}{3} YZ^2 + \left( \rho + \frac{\lambda}{2} \right)Z^2 
\label{eq:Y} \,,\\
\frac{dZ}{d\alpha} &=& Z \left[ 3(X^2 -1) + \frac{1}{2} Y^2
-\rho Y +1 -2X + \frac{1}{3} Z^2 \right]
\label{eq:Z} \ .
\end{eqnarray}
Therefore, we have a 3-dimensional space with a constraint (\ref{constraint}).
 A fixed point in this phase space is defined by $dX/d\alpha=dY/d\alpha=dZ/d\alpha=0$.

First, we seek the isotropic fixed point $X=0$.
From Eq.~(\ref{eq:X}), we see $Z=0$.
The remaining equation (\ref{eq:Y}) yields $Y=-\lambda$ or $Y^2=6$.
The latter solution does not satisfy the constraint (\ref{constraint}). 
Thus, the isotropic fixed point becomes
\begin{eqnarray}
  (X , Y, Z) = ( 0, -\lambda , 0) 
\label{fixedpoint1}\ .
\end{eqnarray} 
This fixed point corresponds to the isotropic power-law solution (\ref{isotropic}).
Indeed, one can check that the solution (\ref{iso-power}) leads to the above fixed point.

Apparently, $Z=0$ and $6 X^2 + Y^2 =6$ give a fixed curve. 
However, this contradicts the constraint (\ref{constraint}).

Now, let us find an anisotropic fixed point.
From Eqs.~(\ref{eq:X}) and (\ref{eq:Y}), we have
\begin{eqnarray}
  Y = \left( 3\rho +\frac{\lambda}{2} \right) X - \lambda \ .
  \label{Y0}
\end{eqnarray}
Eq.~(\ref{eq:X}) gives
\begin{eqnarray}
  Z^2 = - \frac{3X}{X+1} \left[ 3(X^2 -1) + \frac{1}{2} Y^2 \right] \ .
  \label{Z2}
\end{eqnarray}
Using the above results in Eq.~(\ref{eq:Z}), we have
\begin{eqnarray}
  \left( X-2 \right) \left[ \left( \lambda^2 + 8 \rho \lambda + 12 \rho^2 +8 \right)X
              - 2 \left( \lambda^2 +2\rho \lambda -4 \right) \right] =0 \ .
\end{eqnarray}
The solution $X=2$ does not make sense because it implies $Z^2 = -18-36 \rho^2 <0$ by Eqs.~(\ref{Y0}) and (\ref{Z2}).
Thus, an anisotropic fixed point is expressed by
\begin{eqnarray}
 X= \frac{2 \left( \lambda^2 +2\rho \lambda -4 \right)}
            {\lambda^2 + 8 \rho \lambda + 12 \rho^2 +8} \ .
\end{eqnarray}
Substituting this result into Eq.~(\ref{Y0}), we obtain 
\begin{eqnarray}
 Y= - \frac{12 \left( \lambda +2\rho  \right)}
            {\lambda^2 + 8 \rho \lambda + 12 \rho^2 +8} \ .
\end{eqnarray}
Eq.~(\ref{Z2}) yields
\begin{eqnarray}
 Z^2 =  \frac{ 18 \left( \lambda^2 +2\rho \lambda -4 \right)
             \left(-\lambda^2 + 4\rho \lambda +12 \rho^2 +8\right) }
            {\left( \lambda^2 + 8 \rho \lambda + 12 \rho^2 +8 \right)^2}  \ .  
\end{eqnarray}
Note that from the last equation, we find that $\lambda^2 +2\rho \lambda > 4$ is 
required for this fixed point to exist under inflation $\lambda \ll \rho$. 
It is not difficult to confirm that
this fixed point corresponds to the anisotropic power-law solution (\ref{anisotropic}).

Next, we examine the linear stability of the fixed points.
The linearized equations for Eqs.~(\ref{eq:X}), (\ref{eq:Y}), (\ref{eq:Z}) 
are given by
\begin{eqnarray}
\frac{d\delta X}{d\alpha} &=& 
  \left( \frac{1}{3}Z^2 + 9 X^2 + \frac{1}{2} Y^2 -3 \right) \delta X
  + XY \delta Y + \frac{2}{3} \left( X+1 \right)Z \delta Z \,,\\ 
\frac{d\delta Y}{d\alpha} &=& 6X\left( Y + \lambda \right) \delta X 
 + \left\{ 3\left( X^2 -1 \right) 
    + \frac{1}{2}Y^2 +Y\left( Y+\lambda \right) 
                + \frac{1}{3} Z^2 \right\} \delta Y \nonumber\\ 
    &&  \qquad          + \left( \frac{2}{3}Y +2\rho + \lambda \right) Z\delta Z 
\,,\\
\frac{d\delta Z}{d\alpha} &=& 
2(3X -1)Z\delta X +\left( Y-\rho \right)Z \delta Y \nonumber\\
&&   \quad +\left(3X^2 + \frac{1}{2} Y^2 + Z^2
-2X -\rho Y -2 \right)\delta Z\,.
\end{eqnarray}
In the case of the isotropic fixed point Eq.~(\ref{fixedpoint1}), 
these equations reduce to
\begin{eqnarray}
 \frac{d\delta X}{d\alpha} &=& 
  \left( \frac{1}{2} \lambda^2 -3 \right) \delta X \,,\\ 
\frac{d\delta Y}{d\alpha} &=&  
  \left(  \frac{1}{2} \lambda^2 -3 \right) \delta Y  \,,\\
\frac{d\delta Z}{d\alpha} &=& 
  \left[ \frac{1}{2}\lambda^2 -2 + \rho \lambda \right]\delta Z\,.
\end{eqnarray}
We see that the coefficient in the right hand side of above equations becomes 
negative when $\lambda^2 +2\rho \lambda < 4$ during inflation
$\lambda \ll 1$, which means 
the isotropic fixed point is an attractor under these conditions
and the isotropic fixed point becomes stable in this parameter
region.
In the opposite case, $\lambda^2 +2\rho \lambda > 4$, 
the fixed point becomes a saddle point and unstable. 
In the latter cases, if there exist the vector field whatever small it is,
the vector field destabilize  isotropic inflation.  

\begin{figure}[htbp]
\includegraphics[width=14cm]{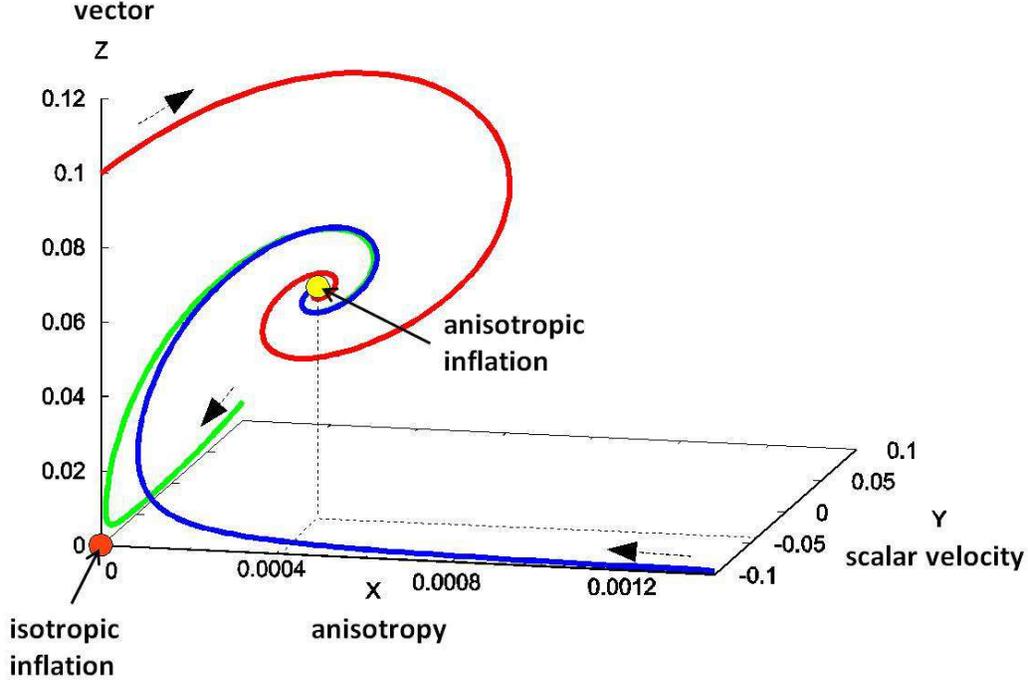}
\caption{The phase flow in $X$-$Y$-$Z$ space is shown for $\lambda = 0.1, \rho=50$. 
 The trajectories converge to the anisotropic fixed point.} 
   \label{fg:flow}
\end{figure}

Now we are interested in the fate of trajectories
around the unstable isotropic fixed point.
We will see that those trajectories converge to an anisotropic fixed point. 
Since we are considering the inflationary universe
$\lambda \ll 1$, the condition $\lambda^2 +2\rho \lambda > 4$
implies $\rho \gg 1$. Under these conditions, we can 
approximately write down the linear equations as
\begin{eqnarray}
 \frac{d\delta X}{d\alpha} &=& 
  -3 \delta X \,,\\ 
\frac{d\delta Y}{d\alpha} &=&  
  -3 \delta Y +  \sqrt{6(\lambda ^2 + 2\rho\lambda- 4)} \delta Z \,,\\
\frac{d\delta Z}{d\alpha} &=& 
  - \frac{1}{2}\sqrt{6(\lambda ^2 +2\rho\lambda - 4) } \delta Y \ .
\end{eqnarray}
The stability can be analyzed by setting 
\begin{eqnarray}
  \delta X = e^{\omega \alpha} \delta \tilde{ X} \ , \hspace{1cm}
   \delta Y = e^{\omega \alpha} \delta \tilde{ Y} \ , \hspace{1cm}
    \delta Z = e^{\omega \alpha} \delta \tilde{ Z} \ .
\end{eqnarray}
Then we find the eigenvalues $\omega$ are given by
\begin{eqnarray}
\omega = -3 \ , 
-\frac{3}{2} \pm i \sqrt{3(\lambda ^2 +2\rho\lambda -4)-\frac{9}{4}} \ . 
\end{eqnarray}
As the eigenvalues have negative real part, 
the anisotropic fixed point is stable.
Thus, the end point of trajectories around the unstable isotropic power-law inflation
must be anisotropic power-law inflation.
In Fig.\ref{fg:flow}, we depicted the phase flow in $X$-$Y$-$Z$ space for $\lambda = 0.1,\ \rho=50$.
We see that the trajectories converge to the anisotropic fixed point indicated by yellow circle.
The isotropic fixed point indicated by orange circle is a saddle point
 which is an attractor only on $Z=0$ plane.
Thus, anisotropic power-law  inflation is an attractor
solution for parameters satisfying $\lambda^2 +2\rho \lambda > 4$~\cite{Kanno:2010nr}.

\section{Anisotropic Inflation: Generality and Universality}
\label{sec:3}

\subsection{Generality}

 Next we want to clarify the generality of anisotropic inflation. 
First, we need to look at the ratio of the shear to the expansion rate $\Sigma/H$ to
 characterize the anisotropy of the inflationary universe. 
 Notice that Eq.(\ref{eq:sigma}) reads 
\begin{eqnarray} 
 \dot{\Sigma} = -3H\Sigma+\frac{2\kappa ^2}{3}\rho _v \ .
\end{eqnarray}
 If the anisotropy converges to a value, i.e. $\dot{\Sigma}$ becomes negligible,
 the terminal value should be given by 
\begin{equation}
\frac{\Sigma}{H} = \frac{2}{3}{\cal R} \ , \qquad {\cal R} \equiv \frac{\rho _v }{ V(\phi)} \ , 
\end{equation}
where  we used the slow roll equation
\begin{eqnarray}
 H^2 = \frac{\kappa ^2}{3} V(\phi)   \ ,
\label{slow-roll:hamiltonian}
\end{eqnarray}
which is derived from Eq.(\ref{eq:hamiltonian}).

In order to realize the above situation, $\rho_v$ must be almost constant.
Assuming that the vector field is subdominant in the evolution equation of 
the inflaton field Eq.(\ref{eq:inflaton}) and conventional single field slow-roll inflation is realized, one can show the coupling function $f(\phi)$ should be
proportional to $e^{-2\alpha}$ to keep $\rho_v$ almost constant. 
 In the slow roll phase, $e$-folding number $\alpha$ is related to the inflaton 
 field $\phi$ as $d\alpha = - \kappa ^2 V(\phi) d\phi / V_\phi$ as usual.  
Then,  the functional form of $f(\phi)$ is determined as
\begin{equation}
f(\phi) = e^{-2\alpha} = e^{2\kappa ^2 \int \frac{V}{V_\phi} d\phi} \ . 
\label{eq:function}
\end{equation}
For the polynomial potential $V\propto \phi ^n$, for example, we have $f=e^{\frac{\kappa ^2 \phi ^2}{n}}$. 
 
The above case is, in a sense, a critical one. What we want to consider is super-critical cases. 
For simplicity, we parameterize $f(\phi)$ by 
\begin{equation}
f(\phi) = e ^{2c\kappa ^2 \int \frac{V}{V_\phi}d\phi},
\label{formula:f}
\end{equation}  
where $c$ is a constant parameter. Now, we look at what happens when $c>1$. 
Note that Eq.(\ref{formula:f}) can be written as
\begin{eqnarray}
  \frac{f_\phi}{f} = 2c\kappa ^2  \frac{V}{V_\phi} \ .
\end{eqnarray}
Then, the condition $c>1$ can be promoted to the condition
\begin{eqnarray}
\frac{1}{2\kappa^2}\frac{f_\phi V_\phi}{f V} >1  \ .
\label{general-condition}
\end{eqnarray}
Thus, any functional pairs $f$ and $V$ which satisfies (\ref{general-condition})
in some range could produce the vector-hair during inflation. 
The equation for the inflaton  becomes
\begin{eqnarray}
\ddot{\phi} = -3\dot{\alpha}\dot{\phi}-V_\phi \left[ 1-\frac{2c}{\epsilon_V}
 {\cal R} \right] \ , \label{eq:inflaton2}
\end{eqnarray}
where we have defined the slow-roll parameter 
\begin{eqnarray}
\epsilon_{V} \equiv \frac{1}{2\kappa ^2} \left( \frac{V_\phi}{ V } \right) ^2 \ .
\end{eqnarray}
 In this case, if the vector field is initially small 
 ${\cal R} \ll \epsilon_V /2c $, then the conventional single field slow-roll inflation is realized. During this stage $f\propto e^{-2c\alpha}$
  and the vector field grows as $\rho_v \propto e^{4(c-1)\alpha}$. Therefore, the vector
  field eventually becomes relevant to the inflaton dynamics Eq.(\ref{eq:inflaton2}). 
  Nevertheless, the accelerating expansion of the universe will continue. 
 The point is that  ${\cal R}$ cannot exceed $\epsilon_{V}/2c$.
 In fact, if ${\cal R}$ exceeds $\epsilon_{V}/2c$, the inflaton field $\phi$
 does not roll down, which makes 
 $\rho _v = p_A^2 f(\phi)^{-2}e^{-4\alpha-4\sigma} /2$
 decrease. Hence,  $\rho_v \ll V(\phi)$ always holds. 
 In this way, there appears an attractor where
 the inflation continues even when the vector field affects the inflaton dynamics. 

The inflaton dynamics is determined by solving the slow-roll equation:
\begin{equation}
 -3\dot{\alpha}\dot{\phi}-V_{\phi}+p_A^2f^{-3}f_{\phi}e^{-4\alpha -4\sigma} =0 \ . 
 \label{eq:inflaton3}
 \end{equation}
Using the slow-roll equation (\ref{slow-roll:hamiltonian}), this yields
\begin{equation}
\frac{d \phi}{d \alpha}=\frac{\dot{\phi}}{\dot{\alpha}}=-\frac{V_\phi}{\kappa ^2 V} + 2c \frac{p_A^2}{V_\phi}e^{-4\alpha -4\sigma -4c \kappa^2 \int\frac{V}{V_\phi}d\phi} \ .
\label{eq:inflaton4}
\end{equation}
This can be integrated by neglecting the evolutions of $V,V_\phi,\sigma$ as
\begin{eqnarray}
  e^{4\alpha +4\sigma +4c \kappa^2 \int\frac{V}{V_\phi}d\phi}
  = \frac{2c^2 p_A^2}{c-1} \frac{\kappa^2 V}{V_{\phi}^2} 
  \left[ 1+ \Omega e^{-4(c-1)\alpha +4\sigma} \right]  \ ,
\label{exponential}
\end{eqnarray}
where $\Omega$ is a constant of integration. 
Substituting this into the slow-roll equation Eq.(\ref{eq:inflaton4}), we obtain
\begin{eqnarray}
\frac{d\phi}{d\alpha} &=& - \frac{V_{\phi}}{\kappa ^2 V}+\frac{c-1}{c}\frac{V_{\phi}}{\kappa ^2V} \left[ 1+ \Omega e^{-4(c-1)\alpha+4\sigma}   \right] ^{-1} \ .
\end{eqnarray}
Initially $\alpha \rightarrow -\infty $, the second term can be neglected. While, in the future $\alpha \rightarrow \infty$,
the term containing $\Omega$ disappears.  
This clearly shows a transition from the conventional single field slow-roll inflationary phase, where
\begin{eqnarray}
\frac{d\phi }{d\alpha} = - \frac{1}{ \kappa ^2 } \frac{V_{\phi}}{ V}
\label{first-stage}
\end{eqnarray} 
holds, to what we refer to as the second inflationary phase, where the vector field is relevant to the inflaton dynamics and the inflaton gets $1/c$ times slower as 
\begin{eqnarray}
\frac{d\phi }{ d\alpha} = - \frac{1}{c} \frac{1}{ \kappa^2} \frac{V_{\phi} }{ V} \ .
\label{second-stage}
\end{eqnarray}
 In the second inflationary phase, we can use the formula (\ref{exponential}) dicarding $\Omega$ term 
and rewrite the energy density of the vector field as
\begin{equation}
\rho _v =\frac{p_A^2}{2} e^{-4\alpha-4\sigma-4c \kappa^2 \int\frac{V}{V_\phi}d\phi}
 = \frac{1}{2} \frac{c-1}{c^2} \epsilon_{V} V(\phi) \ ,
\end{equation}
which yields the anisotropy 
\begin{eqnarray}
\frac{\Sigma}{H} = \frac{2}{3}{\cal R} = \frac{1}{3} \frac{c-1}{c^2} \epsilon_{V} \ .
\end{eqnarray}
 Moreover, from Eqs.(\ref{eq:hamiltonian}) and (\ref{eq:alpha}), 
 the slow-roll parameter defined in terms of the scale factor becomes
\begin{equation}
\epsilon _{H}\equiv -\frac{\ddot{\alpha}}{\dot{\alpha}^2}=-\frac{1}{\dot{\alpha}^2} \left(-\frac{1}{2} \kappa ^2 \dot{\phi}^2-\frac{2}{3}\kappa ^2 \rho_v \right) = \frac{1}{c} \epsilon_{V} \ ,
\end{equation}
where we neglected the anisotropy and used relations (\ref{slow-roll:hamiltonian}) and (\ref{second-stage}).
Thus we have a remarkable result~\cite{Watanabe:2009ct}
\begin{equation}
\frac{\Sigma}{H} = \frac{1}{3}\frac{c-1}{c}\epsilon_{H}.
\end{equation}
Therefore, for a broad class of potential and gauge kinetic functions, there exist
anisotropic inflationary solutions.

\subsection{Example: Chaotic Inflation}

In order to make the statement concrete, we consider chaotic inflation with
the potential 
\begin{eqnarray}
 V(\phi ) = \frac{1}{2} m^2 \phi^2  \ , 
\end{eqnarray}
where $m$ is the mass of the inflaton.
For this potential, the coupling function becomes $f(\phi)=e^{c \kappa^2\phi^2 /2}$. 
It is instructive to see what happens by solving
  Eqs.(\ref{eq:hamiltonian})-(\ref{eq:inflaton}) numerically~\cite{Watanabe:2009ct}.
 \begin{figure}[ht]
\includegraphics[width=12cm]{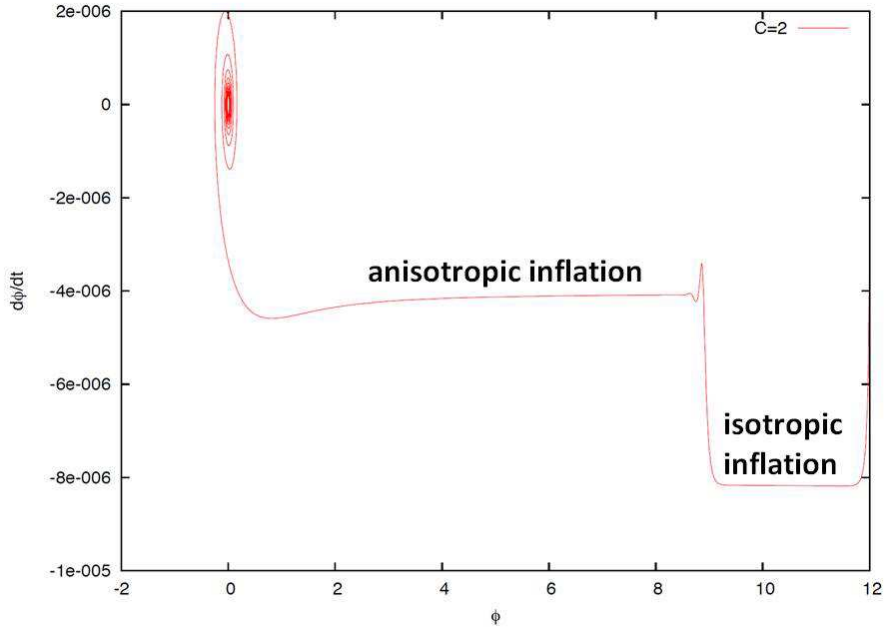}
\caption{Phase flow for $\phi$ is shown. 
 Here, we took the parameters $c=2$ and 
  $\kappa m=10^{-5} $. We also took initial conditions 
 $\phi_i=12$ and $\dot{\phi}_i=0$.
 There are two different slow-roll phases.
 The transition occurs around $\kappa\phi= 9$.}
\label{fg:phase}
\end{figure}
\begin{figure}[h]
\includegraphics[width=12cm]{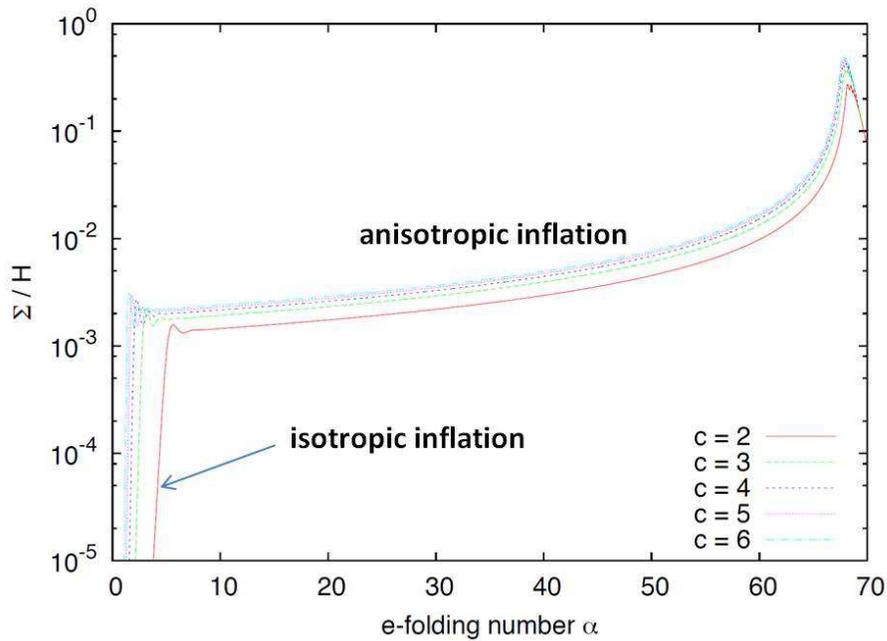}
\caption{
Evolutions of the anisotropy $\Sigma/H$ for various $c$ 
with respect to the $e$-folding number are shown.
One can see the attractor behavior of the anisotropy. }
\label{fg:ce-ratio}
\end{figure}

 In Fig. \ref{fg:phase}, we have shown the phase flow
  in $\phi-\dot{\phi}$ space 
where we can see two slow-roll phases. The first one is the 
  conventional inflationary phase  and the second one is the anisotropic inflationary phase.
As usual, inflation ends with oscillation around the bottom of the potential. 
This tells us that isotropic inflation corresponds to a saddle point
and ansiotropic inflation is an attractor in the slow roll phase.
In contrast to anisotropic power-law inflation, the above chaotic anisotropic inflation
 is a transient phase in the whole phase space.

 In Fig.\ref{fg:ce-ratio},
 we have calculated the evolution of the anisotropy 
 $\Sigma/H \equiv \dot{\sigma}/\dot{\alpha}$ for various parameters $c$
 under the initial conditions $\sqrt{c}\kappa\phi_i=17$. 
As expected, all of solutions show a rapid growth of anisotropy
in the first slow-roll phase which corresponds to the conventional inflation.
However, the growth of the anisotropy eventually stops at the order of the slow roll parameter.
Notice that this attractor like behavior is not so sensitive to the parameter $c$.
As one can see there is a sufficient amount of e-folding number during an anisotropic inflation.

\subsection{Universality}

As we have seen, the anisotropy satisfies the inequality
\begin{eqnarray}
   \frac{\Sigma}{H} \leq  \epsilon_H \ .
\end{eqnarray}
This inequality holds universally for any potential functions. 
This result is reasonable because the cosmic no-hair theorem holds 
for a gravity system with a positive cosmological constant dominating the universe. 
Since the deviation from the exact
de Sitter is characterized by the slow roll parameter, the deviation from the
isotropic expansion must be proportional to the slow roll parameter.

\subsection{Variety of Models}

There are many models which realize anisotropic inflation.
We can generalize single field inflation to multi-field inflation models~\cite{Emami:2010rm,914565}.
Indeed, in almost all kind of models including small field and hybrid inflation, there exists
anisotropic inflation. Actually, a more wide range of anisotropic inflationary
models are discussed in ~\cite{Dimopoulos:2010xq}. 
We can extend the standard kinetic term to the Born-Infeld type~\cite{Moniz:2010cm,arXiv:1105.4455,Do:2011zz}. 
In this direction, we may find a stringy realization of anisotropic inflation.
We can introduce a mass term to the vector field, that is, a vector curvaton~\cite{Dimopoulos:2006ms,arXiv:1107.2779}. 
It is interesting to study cosmological consequences of the vector curvaton scenario in detail.
It is also possible to extend the model to non-abelian gauge fields~\cite{arXiv:1103.6164}.
In this case, we have  more complicated dynamics which would lead to interesting phenomenology.
Interestingly, there are other non-abelian gauge field models~\cite{arXiv:1102.1513,arXiv:1102.1932,arXiv:1109.5573}.
It is interesting to extend the analysis in this paper to other Bianchi type models~\cite{arXiv:1109.3456}.
We can consider inflation with multi-vector fields~\cite{Yamamoto:2012tq} and other tensor fields
to realize anisotropic inflation.

\section{Statistical Anisotropy in Primordial Fluctuations}
\label{sec:4}

So far, we have shown the existence of the anisotropic hair in a variety of inflationary models.
It is interesting to see how to test anisotropic inflation using observations.
To this end, we need the information of fluctuations in anisotropic inflation.
Since the background is anisotropically expanding, we cannot use
the standard cosmological perturbation theory~\cite{Tomita:1985me,Noh:1987vk,Dunsby:1993fg,Gumrukcuoglu:2006xj,
Pereira:2007yy,Gumrukcuoglu:2007bx,Gumrukcuoglu:2008gi,Pitrou:2008gk}.
In this section, we classify perturbations under the 
2-dimensional rotational symmetry and obtain the quadratic actions
for 2-dimensional scalar and vector sectors. 
In order to grasp the meaning of variables, we start with the isotropic case
and make a gauge transformation
from the flat slicing gauge to the appropriate gauge for 2-dimensional
classification. Then, the resultant gauge can be promoted to the anisotropic 
spacetime. The gauge we have chosen makes the analysis and the interpretation 
easier. Once the gauge is fixed, it is straightforward to calculate the quadratic action.

\subsection{Gauge Fixing and Classification of perturbations}

First, we start with the spatially homogeneous and isotropic universe.
For simplicity, we consider flat space. 
\begin{eqnarray}
ds^2 = a^2 (\eta) \left[ -d\eta^2 + \delta_{ij} dx^i dx^j \right] \ ,
\end{eqnarray}
where we took a conformal time $\eta$. 
In that case, we can use 3-dimensional rotational symmetry to classify
the perturbed metric. 
When we want to have the diagonal quadratic action, we take the following gauge
\begin{eqnarray}
ds^2 = a^2 \left[ -(1+2A)d\eta^2 +2 (B_{,i}+V_i ) d\eta dx^i
+  (\delta_{ij} + h_{ij} ) dx^i dx^j \right] \ ,
\end{eqnarray}
where we imposed $V_{i,i}=0$ and $ h_{ij,j}=h_{ii}=0$.
If we ignore vector and tensor perturbations $V_i , h_{ij}$,
the above gauge is called the flat slicing gauge.
Now, let us move on to the Fourier space. Since there exists 3-dimensional
rotation symmetry, we can take a wavenumber vector to be
${\bf k} = (k,0,0)$. Then, the perturbed metric has the following components:
\begin{equation}
\delta g_{\mu \nu} =\left(
\begin{array}{ccccc}
& -2 a^2 A~ &  ~a^2 B_{,x}~ & ~a^2 V_2~ & ~a^2 V_3~
\vspace{1mm}\\\vspace{1mm}
& \ast  & 0 &    0            &   0              
\\\vspace{1mm}
& \ast & \ast & a^2 h_{+}      &  -a^2 h_{\times} 
\\
& \ast     & \ast & \ast &  -a^2 h_{+}
\end{array}
\right) \ . \hspace{5mm} * {\rm ~is ~symmetric ~part}.
\end{equation}
Here, we utilized the special choice ${\bf k} = (k,0,0)$ to solve
the constraints $V_{i,i}=0$ and $ h_{ij,j}=h_{ii}=0$. With the same 
reason, only $B_{,x}$ remains. We defined $h_{yz}=-h_{\times}, 
h_{yy}=-h_{zz}=h_{+}$.
Now, we will pretend that we have only 2-dimensional rotation symmetry 
in $y$-$z$ plane. In that case, at best,
we can take ${\bf k}=(k_x, k_y ,0)$. Hence, 
we make a rotation in the $x-y$-plane so that the wavenumber vector becomes
${\bf k}=(k_x, k_y ,0)$. 
\begin{eqnarray}
  \left(
\begin{array}{cc}
& k_x \\
& k_y \\
& 0  
\end{array}
\right)
= \frac{1}{k}\left(
\begin{array}{cccc}
& k_x~ & ~-k_y~ & ~0~ \\\vspace{1mm}
& k_y & k_x & 0 \\
& 0 & 0 & 0
\end{array}
\right)
 \left(
\begin{array}{cc}
& k \\
& 0 \\
& 0   
\end{array}
\right)  
\ ,
\end{eqnarray}
where we have a relation $k^2 = k_x^2 + k_y^2$.
Under this rotation, the perturbed metric becomes
\begin{equation}
\hspace{-2cm} \delta g_{\mu \nu} =\left(
\begin{array}{ccccc}
& -2 a^2 A~ & ~\frac{k_x}{k} a^2 B_{,x} -\frac{k_y}{k} a^2 V_2~ 
& ~\frac{k_y}{k} a^2 B_{,x} +\frac{k_x}{k} a^2 V_2~          
& a^2 V_3~ 
\vspace{2mm}\\\vspace{2mm}
& \ast &  a^2 \frac{k_y^2}{k^2}h_{+} 
& -a^2 \frac{k_x k_y}{k^2}  h_{+} & a^2 \frac{k_y}{k} h_{\times}  
\\\vspace{2mm}
& \ast & \ast    
& a^2 \frac{k_x^2}{k^2} h_{+} &  -a^2 \frac{k_x}{k} h_{\times} 
\\
& \ast & \ast & \ast &  -a^2 h_{+}
\end{array}
\right) \ .
\end{equation}
To simplify the perturbations, we can make use of gauge transformation
\begin{eqnarray}
\delta g_{\mu\nu} \rightarrow 
\delta g_{\mu\nu} + \xi_{\mu ; \nu} +\xi_{\nu;\mu} \ ,
\end{eqnarray}
where the semicolon denotes the covariant derivative with respect to the background
metric. Taking the parameter 
$$
\xi^0 =0 \ , \quad \xi^x = \frac{k_x}{2i k^2 }h_{+} \ , \quad
\xi^y = \frac{k_y}{2i k^2} h_{+} \ , \quad
\xi^z = \frac{k_x}{ik_y k } h_{\times} \ ,
$$
we obtain
\begin{equation}
\hspace{-2cm} \delta g_{\mu \nu} =\left(
\begin{array}{ccccc}
& -2 a^2 A~ 
& ~\frac{k_x}{k}a^2B_{,x}+\cdots ~ 
& ~\frac{k_y}{k}a^2 B_{,x}+\cdots ~ 
& ~a^2 V_3+\cdots ~  
\vspace{2mm}\\\vspace{2mm}
& \ast &  a^2 h_{+}  
&  0  &   a^2 \frac{k}{k_y} h_{\times}  
\\\vspace{2mm}
& \ast &  \ast    
&  a^2  h_{+}   &  0 
\\
& \ast & \ast 
& \ast &  -a^2 h_{+}
\end{array}
\right) \ ,
\label{gauge-A}
\end{equation}
where we have omitted some unimportant parts.
It should be noted that we did not change slicing but
 performed only the spatial coordinate transformation. Therefore, 
 we are still working in the flat slicing where the 3-dimensional scalar curvature
 vanishes. 
 
In our anisotropic inflation models, the available symmetry is actually small. 
The background metric is given by
\begin{equation}
ds^2_b =a(\eta) ^2(-d\eta ^2+dx^2)+b(\eta) ^2(dy^2+dz^2),
\end{equation}
that is, $ a=e^{\alpha -2\sigma}, b=e^{\alpha +\sigma}, d\eta=dt/a$.
Notice that the conformal time in anisotropic inflation is the conformal
time in 2-dimensional part $(\eta , x)$. 
Even in this anisotropic spacetime,
as we have done in (\ref{gauge-A}), one can take the following gauge
\begin{equation}
\delta g_{\mu \nu} =\left(
\begin{array}{ccccc}
& \delta g_{00}~ & ~\delta g_{0x}~ & ~\delta g_{0y}~ & ~\delta g_{0z}~
\vspace{1mm}\\\vspace{1mm}
& *  &  \delta g_{xx}  &  0 &  \delta g_{xz} 
\\\vspace{1mm}
&  * &  * &  \delta g_{yy}  &  0 
\\\vspace{1mm}
&  *  &  *   & * &  \delta g_{zz}
\end{array}
\right) \ ,
\end{equation}
where we can impose further conditions so that the perturbed metric goes back
to (\ref{gauge-A}) in the isotropic limit. 

One can classify the perturbed metric using the rotational symmetry in $y-z$-plane.
In 2-dimensional flat space, an arbitrary vector $m^a $ where $a=y,z$ can be 
decomposed into the scalar part $m^a_{,a}\neq 0$ and the vector part $m^a_{,a} =0$. 
Since there exists no tensor part in 2-dimensions, 2-dimensional tensor can be
constructed from the 2-dimensional vector. 
Because of the symmetry, the scalar and vector parts are not mixed in the equations.
Thus, the metric perturbations
can be classified into the scalar sector and the vector sector. 
Thanks to the symmetry in the $y-z$ plane,  without loss of generality, 
 we can take the wavenumber vector to be ${\bf k} = (k_x , k_y ,0)$. 
Hence, the vector sector in 2-dimensional classification can be identified as
$\delta g_{0z} , \delta g_{xz}$ in the above perturbed metric.
The remaining components $\delta g_{00} , \delta g_{0x} , \delta g_{0y} ,
\delta g_{xx}, \delta g_{yy}, \delta g_{zz} $ belong to the scalar sector.

\subsubsection{2d vector sector}

 Thus, the perturbations that belong to 2d vector perturbations, 
 can be written as
\begin{equation}
\delta g_{\mu \nu}^{\rm vector} =\left(
\begin{array}{ccccc}
& 0~ & ~0~ & ~0~ & ~b^2 \beta_3~ \vspace{1mm}\\\vspace{1mm}
& *  &  0  &  0 & b^2 \Gamma \\\vspace{1mm}
&  * &  * &  0  &  0 \\
&  *  &  *   & * &  0
\end{array}
\right) \ ,
\end{equation}
where we have incorporated the anisotropy while keeping the spatial 
scalar curvature to be zero. 
As to the vector field, we can take 
\begin{eqnarray}
\delta A_{\mu}^{\rm vector} = \left(0 \ , 0 \ , 0 \ ,  D \right) \ .
\end{eqnarray}
Note that we have no residual gauge transformation and, in particular,
 $D$ is a gauge invariant under abelian gauge transformations. 
 And, as we have seen in (\ref{gauge-A}), $\Gamma$ corresponds to
  the cross-mode polarization of gravitational waves in the isotropic limit $a=b$.

Using this gauge, we can calculate the quadratic action as
\begin{eqnarray}
\hspace{-2.5cm} S^{{\rm vector }}
\hspace{-1.5cm} &=&
\int d\eta d^3 x \left[~ 
\frac{b^4}{4a^2} \beta^2_{3,x} + \frac{b^2}{4} \beta^2_{3,y} 
- \frac{b^4}{2a^2} \Gamma' \beta_{3,x} 
+ \frac{f^2 v' b^2}{a^2} \beta_3 D_{,x}     \right. \nonumber \\
\hspace{-1.5cm}&&  \left. \qquad\qquad
-\frac{b^2}{4} \Gamma^2_{,y} + \frac{b^4}{4a^2} \Gamma^{\prime 2} 
-\frac{f^2a^2}{2b^2} D_{,y}^2
-\frac{1}{2}f^2 D_{,x}^{2}+\frac{f^2}{2} D^{\prime 2}
-\frac{f^2 v' b^2}{a^2} D' \Gamma ~\right] \ .  
\end{eqnarray}
Since the perturbed shift function $\beta_3$ does not have a time derivative,
  it is not dynamical. 
There are two physical degrees of freedom $\Gamma$ and $D$ in this
2-dimensional  vector sector. 

\subsubsection{2d scalar sector}

For the 2-dimensional scalar sector, we define the metric perturbations
\begin{equation}
\delta g_{\mu \nu}^{\rm scalar} =\left(
\begin{array}{ccccc}
& -2 a^2 \Phi~ & ~a \beta_1~  & ~a \beta_2~ & ~0~ \vspace{2mm}\\\vspace{1mm}
& *  &  2 a^2 G  &  0 & 0 \\\vspace{1mm}
&  * &  * &  2 b^2 G  &  0 \\
&  *  &  *   & * & -2 b^2 G
\end{array}
\right) \ ,
\end{equation}
where we have kept the spatial scalar curvature vanishing. 
The scalar perturbation will be represented by $\delta \phi$. 
The variable $G$ and $\delta \phi$ are the gauge invariant variables 
that correspond to the plus mode of gravitational waves 
and the scalar perturbations, respectively, in the isotropic  limit $a=b$. 
And, we set the perturbed vector to be
\begin{eqnarray}
\delta A^{\rm scalar}_{\mu} = \left( \delta A_0 \ , 0 \ ,   J \ , 0 \right)  \ ,
\end{eqnarray} 
where we have fixed the abelian gauge by putting the longitudinal component 
to be zero.
From these ansatz, we can calculate the quadratic action as
\begin{eqnarray}
\hspace{-2.5cm} S^{{\rm scalar}}
  &=& \int d^3 x d\eta 
 \left[~ 
 \frac{b^2}{2a^2} f^2 \delta A_{0,x}^2 + \frac{f^2}{2} \delta A_{0,y}^2
 +\frac{b^2}{a^2} f^2 v' \left(G+\Phi \right) \delta A_{0,x} -f^2 J' \delta A_{0,y}
   \nonumber \right. \\ 
\hspace{-2cm}&& -2 \frac{b^2}{a^2} ff_\phi v'\delta \phi \delta A_{0,x}
+\frac{1}{4} \beta_{1,y}^2
 -\frac{1}{2} \beta_{2,x} \beta_{1,y} + 2\frac{bb'}{a} \Phi_{,x} \beta_1 
 - \frac{b^2}{a} \phi' \delta\phi_{,x} \beta_1 + \frac{1}{4} \beta_{2,x}^2 
   \nonumber \\
\hspace{-2cm}&& + a \left( \frac{a'}{a}+\frac{b'}{b} \right) \beta_2 \Phi_{,y}
-a \left(\frac{a'}{a} -\frac{b'}{b} \right) \beta_2 G_{,y} 
 +  \frac{f^2}{a} v' \beta_2 J_{,x} -a \phi' \beta_2 \delta\phi_{,y} 
  + \frac{1}{2} f^2 J^{\prime 2}  
   \nonumber\\
\hspace{-2cm}&&  -\frac{1}{2} f^2 J_{,x}^2 
   +b^2 G^{\prime 2} -a^2 G_{,y}^2 -b^2 G_{,x}^2
+\frac{1}{2}b^2 \delta\phi^{\prime 2}
 -\frac{a^2}{2} \delta\phi_{,y}^2 - \frac{b^2}{2} \delta\phi_{,x}^2
 -\frac{1}{2} a^2 b^2 V_{\phi\phi} \delta\phi^2 
 \nonumber \\
\hspace{-2cm}&&  
  + \frac{b^2}{2a^2} \left( f_\phi^2 +ff_{\phi\phi} \right) 
 v^{\prime 2} \delta\phi^2          
 - a^2 b^2 V  \Phi^2 +  \frac{b^2}{2 a^2} f^2 v^{\prime 2}  G^2 
 - 2a^2 b^2 V \Phi G    -2bb' \Phi' G   \nonumber \\
\hspace{-2cm}&&  \left.  \qquad \qquad
 - \left( \frac{b^2}{a^2} ff_\phi v^{\prime 2} + a^2 b^2 V_\phi \right)
  \delta\phi \left( G+\Phi \right) +b^2 \phi' \delta\phi' \left( G-\Phi \right)
  ~\right]    \ .
\end{eqnarray}
Here, $S^{{\rm scalar}}$ consists of $\Phi, \beta_1, \beta_2, G, \delta A_0 , \delta \phi$
 and $J $. Among them, $\Phi , \beta_1, \beta_2$ and $\delta A_0 $ 
 are non-dynamical and can be eliminated.

In order to calculate the statistical properties
of primordial fluctuations from anisotropic inflation
~\cite{Himmetoglu:2009mk}\cite{Dulaney:2010sq}\cite{Gumrukcuoglu:2010yc}\cite{Watanabe:2010fh},
we need to reduce the action to the one for physical variables.
Then, we can quantize the system and specify the vacuum state.
We analyze the vector sector and the scalar sector, separately. 


\subsection{Action in slow roll approximation}

First, let us consider the vector sector and eliminate non-dynamical variable $\beta_3$
 from the action for the 2-dimensional vector sector.
Now, we define canonically normalized variables as
\begin{eqnarray}
\bar{\Gamma}  \equiv  \frac{b |k_y|}{\sqrt{2}k}\Gamma, \qquad
\bar{D}  \equiv  fD.
\end{eqnarray}
Then, using these canonical variables, 
we obtain the reduced action for physical variables
\begin{eqnarray}
\hspace{-2cm}S^{\rm vector} &=& \int d\eta d^3 k \left[ 
\frac{1}{2} |\bar{\Gamma}^{'}|^2
+\frac{1}{2}\left( \frac{(b/k)^{''}}{(b/k)}-k^2 \right) |\bar{\Gamma}|^2 
\right. \nonumber\\
\hspace{-2cm} &&\qquad \qquad \qquad \left.
 +\frac{1}{2}|\bar{D}^{'}|^2
 +\frac{1}{2}\left( \frac{f^{''}}{f}-k^2
 -2\frac{f^2v^{'2}}{a^2}\frac{k_x^2}{k^2} 
                         \right) |\bar{D}|^2  \right. \nonumber\\
\hspace{-2cm} &&\qquad \qquad \quad \left. 
 + \frac{1}{\sqrt{2}} \frac{fv^{'}}{a}\frac{a}{b}\frac{k_y}{k} 
 \left\{ \bar{\Gamma}^{'}\bar{D}^{*} +  \bar{\Gamma}^{*'}\bar{D}
 +\frac{(k/b)^{'}}{(k/b)}\left( \bar{\Gamma}\bar{D}^{*} + \bar{\Gamma}^{*} \bar{D} \right)
                              \right\} \right]  \ ,
\label{vec-action}
\end{eqnarray}
where $k$ is time dependent and given by 
\begin{eqnarray}
k(\eta) \equiv \sqrt{k_x^2+ \frac{a^2 (\eta)}{b^2 (\eta )} k_y^2} \ ,
\end{eqnarray}
which becomes constant in the isotropic limit $a=b$.
In the isotropic limit $a=b$,
 $\bar{\Gamma} $ and $\bar{D}$ represent the cross-mode of gravitational waves
and vector waves, respectively. The second line in the action (\ref{vec-action})
describes how both waves are interacting to each other. 

Next, we use the slow roll approximation to simplify the action.
To obtain the homogeneous background metric, 
we integrate the following equations
\begin{eqnarray}
-\frac{\dot{H}}{H^2} = \epsilon _H, \qquad
\frac{\Sigma}{H}  = \frac{1}{3} I \epsilon _H \ ,
\end{eqnarray}
by assuming $\epsilon_H^{'}/\epsilon_H \ll a^{'} /a$.
The resultant expressions are 
\begin{eqnarray}
a = (-\eta )^{-1-\epsilon _H }, \qquad
b = (-\eta )^{-1-\epsilon _H - I \epsilon_H } \ .
\end{eqnarray}
In this approximation, the universe shows anisotropic power law inflation.
We should recall, in the second inflationary phase, the variable $I$ is given by
\begin{equation}
I = \frac{c-1}{c} \ .
\end{equation}
Note that the range $(1,\infty )$ for $c$ corresponds to $(0,1)$ for $I$.
Using the definition of ${\cal R}$, we obtain
\begin{eqnarray}
\frac{f^2 v^{'2}}{a^2} &=& 3(-\eta )^{-2} I \epsilon _H \ . 
\end{eqnarray}
From Eq.~(\ref{eq:Ax}), the background equation for the vector can be found as
\begin{eqnarray}
\left[ \frac{f^2v^{'}b^2}{a^2} \right] ^{'} = 0  \ .
\end{eqnarray}
From this equation, it is easy to deduce the relation
\begin{eqnarray}
\frac{f^{'}}{f} = (-\eta )^{-1}\left[ -2 -3\epsilon _H + \eta _H 
-2 I \epsilon _H \right] \ ,
\end{eqnarray}
where $\eta _H$ is a slow-roll parameter defined by
\begin{equation}
\frac{\epsilon _H ^{'}}{\epsilon _H} 
= 2 \frac{(e^{\alpha})^{'}}{e^{\alpha}} \left( 2 \epsilon _H - \eta _H \right) 
= 2 (2 \epsilon _H -\eta _H) (-\eta )^{-1} \ .
\end{equation}
Of course, $\eta_H$ is not related to the conformal time $\eta$. 
Furthermore, we obtain
\begin{eqnarray}
\frac{f^{''}}{f} &=& 
(-\eta )^{-2} \left[ 2+9\epsilon _H -3 \eta _H + 6 I\epsilon _H \right] \ .
\end{eqnarray}
Substituting these results into the action, we obtain the action
in the slow roll approximation~\cite{Watanabe:2010fh}
\begin{eqnarray}
\hspace{-2.5cm}S^{\rm vector} 
&=& \int d\eta d^3 k \left[
 \frac{1}{2} | \bar{\Gamma}^{'} | ^2 
+\frac{1}{2} \left[ -k^2+ (-\eta )^{-2} \left\{ 2+3\epsilon _H+3I \epsilon _H 
+ 3 I\epsilon _H \sin^2 \theta \right\} \right] |\bar{\Gamma}| ^2
\right. \nonumber \\
 \hspace{-2.5cm}&& \qquad \qquad + \frac{1}{2}|\bar{D}^{'}|^2 
 +\frac{1}{2} \left[ -k^2+ (-\eta)^{-2} \left\{ 2+9\epsilon _H -3\eta _H 
  +6 I\epsilon _H \sin^2 \theta \right\} \right] |\bar{D}|^2 
\nonumber \\
\hspace{-3cm}&& \!\!\!\! \left. + \frac{\sqrt{6I\epsilon_H}}{2}(-\eta)^{-1}
\sin \theta (\bar{\Gamma}^{'}\bar{D}^{*}+\bar{\Gamma}^{*'}\bar{D})
-\frac{\sqrt{6I\epsilon_H}}{2}(-\eta)^{-2}
\sin \theta (\bar{\Gamma}\bar{D}^{*}+\bar{\Gamma}^{*}\bar{D}) \right]  , \quad
\label{eq:lagbegin}
\end{eqnarray}
where we have defined
\begin{eqnarray}
\sin \theta \equiv \frac{k_y a}{ k b} \ .
\end{eqnarray} 
This $\theta$ represents the direction dependence. 
In the isotropic limit $I=0$, 
the Lagrangian for $\bar{\Gamma}$ becomes the familiar one for gravitational 
waves in a Friedman-Lemaitre universe.

In a similar way, we can derive the quadratic action for physical variables
in the 2-dimensional scalar sector. 
Moreover, it is straightforward to deduce the action in the slow roll approximation.
The resultant action is given by~\cite{Watanabe:2010fh}
\begin{eqnarray}
\hspace{-2cm}&&S^{\rm  scalar} = \int d\eta d^3k 
\left[ L^{GG}+L^{JJ}+L^{\phi\phi}+L^{\phi G}
+L^{\phi J}+L^{JG} \right] \ , 
\label{sca-action} 
\end{eqnarray}
where diagonal parts are given by
\begin{eqnarray}
\hspace{-2.5cm}&&L^{GG}=\frac{1}{2}|\bar{G}^{'}|^2 
+\frac{1}{2} \left[-k^2+(-\eta) ^{-2} \left\{ 2+3\epsilon _H+3I\epsilon _H
 + 3 I\epsilon _H \sin^{2} \theta \right\} \right] |\bar{G}|^2,
\label{GG} \\
\hspace{-2.5cm}&&L^{JJ}= \frac{1}{2}|\bar{J}^{'}|^2 
+\frac{1}{2} \left[ -k^2 +(-\eta )^{-2} \left\{ 2+9\epsilon _H -3\eta _H 
          - 6 I\epsilon _H \sin^2 \theta \right\} \right] |\bar{J}|^2, 
\label{JJ} \\
\hspace{-2.5cm}&& L^{\phi \phi} = \frac{1}{2} | \delta\bar{ \phi}^{'}|^2 \nonumber\\
\hspace{-2.5cm}&& +\frac{1}{2} \left[ -k^2 +(-\eta) ^{-2} \left\{ 2 + 9\epsilon _H 
 -\frac{3\eta _H}{1-I}-\frac{12I}{1-I}+\left( 12I \epsilon _H +\frac{24I}{1-I} \right)
  \sin^2 \theta \right\} \right] | \delta\bar{\phi}|^2, 
\label{dphi} 
\end{eqnarray}
and the interaction parts reads
\begin{eqnarray}
\hspace{-2cm}&&L^{\phi G} = -3 I \sqrt{\frac{\epsilon _H}{1-I}} 
(-\eta )^{-2} \sin^2\theta
\left(\bar{G} \delta\bar{\phi}^* +\bar{G}^* \delta\bar{\phi} \right) \ , 
\label{phi:G} \\
\hspace{-2cm}&& L^{\phi J} = \sqrt{ \frac{6I}{1-I}} (-\eta )^{-1} \sin\theta 
\left(\delta\bar{ \phi}^{*'}\bar{J}+\delta\bar{\phi}^{'}\bar{J}^{*}\right) \nonumber\\
\hspace{-2cm}&&  \qquad \qquad  - \sqrt{ \frac{6I}{1-I}}(-\eta )^{-2}\sin \theta
\left(\delta\bar{ \phi}^{*} \bar{J}+\delta\bar{\phi}\bar{J}^*\right) 
\ , 
\label{phi:J} \\
\hspace{-2cm}&& L ^{JG} = -\frac{\sqrt{6I \epsilon _H}}{2} (-\eta ) ^{-1} \sin\theta
\left(\bar{G}^{*'}\bar{J} + \bar{G}^{'} \bar{J}^{*}\right)  \nonumber\\
&& \qquad \qquad    + \frac{\sqrt{6I \epsilon _H}}{2} (-\eta )^{-2}\sin\theta
\left(\bar{G}^{*}\bar{J}+\bar{G}\bar{J}^*\right) \ . 
\label{J:G}
\end{eqnarray}
Here, we defined canonical variables
\begin{equation}
\bar{G}\equiv\sqrt{2}bG \ , \quad  
\bar{J}\equiv \frac{f |k_x| }{k}J \ , \quad  
\delta\bar{ \phi} \equiv b \delta \phi      \ .
\end{equation}
Note that $\bar{G} \ , \bar{J}$ and $\delta\bar{\phi}$ represent
the gravitational waves, the vector waves, and the scalar perturbations, respectively.
The above action shows there exist the interaction among these variables.
We notice the scalar part (\ref{dphi})
contains $I$ without suppression by a slow-roll parameter $\epsilon _H$. 
Therefore, to obtain the quasi-scale invariant spectrum of curvature perturbation, $I$ itself has to be small.

From the actions (\ref{eq:lagbegin}) and (\ref{sca-action}), 
we see there are two sources of statistical anisotropy of fluctuations. 
First, the statistical anisotropy of fluctuations comes from the anisotropic 
expansion itself. Intuitively, this can be understood from the anisotropic 
effective Hawking temperature $H_{\rm eff}/2\pi$, where $H_{\rm eff}$
denotes the effective expansion rate. 
Indeed, the expansion rate in the direction of the background vector is relatively small,
hence the effective Hawking temperature is low.  Then, this direction has less 
fluctuation power compared to the other directions.
Thus, the effective Hawking temperature 
induces the anisotropy in the power spectrum of fluctuations.
This effect is encoded in (\ref{GG}), (\ref{JJ}), and (\ref{dphi}).
The other source of the statistical anisotropy of fluctuations comes from the couplings
(\ref{phi:G}), (\ref{phi:J}) and (\ref{J:G}) due to the background vector field. 
The essential structure of couplings can be understood without complicated calculations.
Take a look at the following term
\begin{eqnarray}
  \sqrt{-g} g^{\mu\alpha} g^{\nu\beta} f^2(\phi) F_{\mu\nu} F_{\alpha\beta} \ .
\end{eqnarray}
Here, we should recall the order of magnitude of background quantities
$$
 \frac{f^2 v^{\prime 2}}{a^2} \sim I \epsilon_H \ , \quad
 \frac{f_\phi}{f} \sim \frac{\kappa^2 V}{V_\phi}
 \sim \frac{1}{\sqrt{\epsilon_H}} \ .
$$
For example, to obtain the $J-G$ coupling, one of $F_{\mu\nu}$ have to be replaced
by the background quantity $v'$. Hence, the coefficients in the $J-G$ coupling
should be proportional to $f v'$ which is of the order of $\sqrt{I\epsilon_H}$. 
This explains the strength of the coupling in (\ref{J:G}). 
Similarly, $J-\delta\phi$ coupling should be proportional to $f_\phi v'$
 because we have to take the variation with respect to $\phi$.
 Hence, we can estimate its magnitude to be $\sqrt{I}$. 
  This explains the interaction term (\ref{phi:J}). Finally, the coupling
$G-\delta\phi$ has a magnitude of the order of $f_\phi v^{\prime 2}$
which is proportional to $I \sqrt{\epsilon_H}$.
This shows a good agreement with the coupling (\ref{phi:G}).
Thus, we can understand why there is a hierarchy among the couplings of
the gravitational waves, the vector waves and the scalar field.

\subsection{Statistical Anisotropy}

In this subsection, we will calculate corrections to power spectrum of various variables
due to the anisotropy. To set the initial conditions, we need to quantize this system 
by promoting canonical variables to operators
which satisfy the canonical commutation relations.
The point is that, with a given wavenumber, the actions (\ref{eq:lagbegin}) 
and (\ref{sca-action}) reduce to 
  those of independent harmonic oscillators in the subhorizon limit 
  $-k\eta \gg 1$.  
We choose the Bunch-Davis vacuum state $|0 \rangle$ 
by imposing the conditions $a_{a,\bf k}|0 \rangle  =0$ at an initial time $\eta _i$.
Here,  $a_{a,{\bf k}}$ is an annihilation operator whose commutation relations are given by
\begin{eqnarray}
\left[ a_{a,\bf k} , a^\dagger _{b\bf k^\prime } \right] 
 = \delta_{ab}\delta ^{(3)}({\bf k-k^\prime}), \qquad
\left[ a_{a, \bf k} , a_{b,\bf k^\prime } \right] 
= 0 \ .
\end{eqnarray}
We are interested in the power spectrum of
 the scalar perturbations
\begin{eqnarray}
\langle 0 \big|\delta \bar{\phi}_{\bf k}(\eta) 
               \delta \bar{\phi}_{\bf p}(\eta) \big| 0 \rangle
\equiv P_{\delta\phi} ({\bf k}) \delta({\bf k} + {\bf p})
    \ ,
\end{eqnarray}
and the power spectrum of  the cross and  plus mode of
gravitational waves
\begin{eqnarray}
&&  \langle 0 \big| \bar{\Gamma}_{\bf k}(\eta) \bar{\Gamma}_{\bf p}(\eta) \big| 0 \rangle
 \equiv P_{\Gamma} ({\bf k}) \delta({\bf k} + {\bf p}) \ , \\
&&  \langle 0 \big| \bar{G}_{\bf k}(\eta) \bar{G}_{\bf p}(\eta) \big| 0 \rangle
 \equiv P_{G} ({\bf k}) \delta({\bf k} + {\bf p})     \ .
\end{eqnarray}
We can also calculate the cross correlation between the plus mode
of gravitational waves and the scalar perturbations
\begin{eqnarray}
\langle 0 \big| \delta \bar{\phi}_{\bf k}(\eta) \bar{G}_{\bf p}  (\eta) \big| 0 \rangle
\equiv P_{\delta\phi G} ({\bf k}) \delta({\bf k} + {\bf p})  \ .
\end{eqnarray}

We treat the anisotropy perturbatively and estimate its magnitude by 
using perturbation in the interaction picture. 
In the interaction picture, 
the expectation value for a physical quantity ${\cal O} (\eta)$ is given by
\begin{equation}
\hspace{-2cm}\langle in \left| {\cal O} (\eta) \right |in \rangle 
= \left< 0 \left| 
\left[ \bar{T}\exp \left( i \int ^{\eta}_{\eta_i} H_I(\eta ^{'}) d\eta ^{'} 
\right) \right] {\cal O} (\eta ) 
\left[ T \exp \left( -i\int ^{\eta}_{\eta_i} H_I(\eta ^{'}) d\eta ^{'} \right) \right] 
\right| 0 \right> \ ,
\end{equation}
where $|in \rangle$ is an in vacuum in the interaction picture, 
 $T$ and $\bar{T}$ denote a time-ordered and an anti-time-ordered product
 and $H_I$ denotes  the interaction part of Hamiltonian
 in this picture. 
 This is equivalent to the following
\begin{eqnarray}
\hspace{-2cm}\langle in \left| {\cal O} (\eta) \right| in \rangle 
&=& \sum _{N=0}^{\infty} i^N \int _{\eta_i}^{\eta} d\eta _N\int _{\eta_i}^{\eta_N} 
d\eta _{N-1} \cdots \int _{\eta_i}^{\eta_2} d\eta _{1} \nonumber \\
&&  \qquad \times \left<0 \left| 
\left[ H_I(\eta_1),\left[H_I(\eta_2), \cdots \left[H_I(\eta _N),{\cal O}(\eta) \right] 
\cdots \right] \right] 
\right| 0 \right>.
\end{eqnarray}
In our analysis, we assume the noninteracting part of Hamiltonian to be that of free fields in deSitter spacetime 
\begin{equation}
L_0 = \sum _{n} \left[ \frac{1}{2} |Q^{'}_n|^2-\frac{1}{2} 
                    \left( k^2 - 2(-\eta )^{-2} \right) |Q_n|^2 \right] \ ,
\end{equation}
and the operators in the interaction picture are given by
\begin{eqnarray}
Q_{n,{\bf k}}(\eta) &=& u(\eta)  a_{n,{\bf k}} 
+ u(\eta)^{*} a_{n,{\bf -k}}^{\dagger}, 
\label{eq:nonint}\\
u(\eta) &\equiv &\sqrt{\frac{1}{2k}}e^{-ik\eta}\left(1-\frac{i}{k\eta} \right) \ ,
\label{eq:modefunc}
\end{eqnarray}
where $Q_n$ represent the physical variables $\bar{D}, \bar{\Gamma}, \bar{G}, \bar{J}, \delta\bar{\phi}$.
And the rest of the Lagrangian (\ref{eq:lagbegin})-(\ref{J:G})
is regarded as the interaction part $L_I = L^{(2)}-L_0$. 
To see the leading effect on the anisotropy in the scalar perturbation, 
which is of the order of $I$, we evaluate the correction due to the interaction
 given by
\begin{eqnarray}
H_I^{\phi J} &\equiv&  \int d^3 k \left[ - L^{\phi J} \right] \nonumber\\
\hspace{-2cm}&=& \int d^3k \left[ -\sqrt{ \frac{6I}{1-I}} (-\eta )^{-1} \sin\theta 
\left(\bar{\delta \phi}^{\dagger'}\bar{J} +\bar{\delta\phi}^{'}\bar{J}^{\dagger}\right)
\right. \nonumber\\
&& \left.  \qquad \qquad  + \sqrt{ \frac{6I}{1-I}}(-\eta )^{-2}\sin \theta
\left(\bar{\delta \phi}^{\dagger} \bar{J} +\bar{\delta\phi} \bar{J}^{\dagger}\right)
\right] 
\ .
\end{eqnarray}
 Note that in the analogy with the slow-roll parameter in the ordinary slow-roll inflation, the term proportional to 
 $I\sin^2 \theta \delta\bar{\phi}
 \delta\bar{\phi}^{\dagger}$ in (\ref{dphi}) can be expected to give the anisotropy  
$\delta \langle in \left| \delta \bar{\phi}_{\bf k}
                         \delta \bar{\phi}_{\bf p} \right| in \rangle 
/ \langle 0 \left| \delta \bar{\phi}_{\bf k} \delta \bar{\phi}_{\bf p} \right| 0 \rangle 
\sim \sin ^2 \theta I N(k) $ 
where $N(k)$ is the $e$-folding number from the horizon exit. 
Thus, the leading correction comes from the interaction
through the term $H^{\phi J}_{I}$. The leading correction is given by
\begin{eqnarray}
\hspace{-1cm} &&\delta \langle in \left| \delta \bar{\phi}_{\bf k} (\eta) 
                   \delta \bar{\phi}_{\bf p} (\eta) \right| in \rangle  \nonumber \\
\hspace{-1cm}&& \qquad = i^2 \int_{\eta _i}^{\eta}d\eta_2\int_{\eta _i}^{\eta _2} d\eta _1 
\left< 0 \left|
 \left[ H^{\phi J}_I(\eta_1),\left[H^{\phi J}_I(\eta_2),\delta \bar{\phi}_{\bf k} (\eta)
 \delta \bar{\phi}_{\bf p} (\eta) \right] \right] 
 \right| 0 \right> \ .  \ 
\end{eqnarray}
Using Eqs.(\ref{eq:nonint}) and commutation relations for the creation and annihilation operators, we obtain the anisotropy expressed as follows
\begin{eqnarray}
\hspace{-2cm}&&\frac{\delta \langle in \left| \delta \bar{\phi}_{\bf k}
                             \delta \bar{\phi}_{\bf p} \right| in \rangle}
{\langle 0 \left| \delta \bar{\phi}_{\bf k} 
                 \delta \bar{\phi}_{\bf p} \right| 0 \rangle} (\eta) \nonumber\\
\hspace{-2cm}&=& \frac{24 I}{1-I} \sin ^2 \theta 
 \int ^{\eta} _{\eta_i} d\eta _2 \int ^{\eta_2}_{\eta_i} d\eta_1 
  \frac{8}{|u(\eta)|^2} {\rm Im}
\left[  -(-\eta_2)^{-1}  u^{'}(\eta_2) u^{*}(\eta) 
   + (-\eta _2 )^{-2} u(\eta_2 )u^{*}(\eta )  \right] \nonumber\\
\hspace{-2cm}&\ &\, \times {\rm Im}
\left[ u(\eta_1 )u^{*}(\eta _2)\left\{ -(-\eta_ 1)^{-1} u^{'}(\eta _1)u^{*}(\eta ) 
 + (-\eta_1)^{-2} u(\eta_1 )u^{*}(\eta) \right\}  \right] \ ,
\end{eqnarray} 
where ${\rm Im}$ denotes the imaginary part.
Substituting the function form of $u$ (\ref{eq:modefunc}) 
and introducing time variables $\chi \equiv k\eta$, $\chi_1 \equiv k\eta_1$ and
$\chi_2 \equiv k\eta_2$, we have
\begin{eqnarray}
\hspace{-2cm}&&\frac{\delta \langle in \left| \delta \bar{\phi}_{\bf k}
                             \delta \bar{\phi}_{\bf p} \right| in \rangle}
{\langle 0 \left| \delta \bar{\phi}_{\bf k}
                             \delta \bar{\phi}_{\bf p} \right| 0 \rangle} (\chi) \nonumber\\
\hspace{-2cm}&=& \frac{6I}{1-I} \sin ^2 \theta \int ^{\chi}_{\chi_i}d\chi_2 \int ^{\chi_2}_{\chi_i} d\chi_1 
 \frac{8}{1+\frac{1}{(-\chi)^2}}\frac{1}{-\chi_1}\frac{1}{-\chi_2}\left[ \cos(-\chi_2+\chi)-\sin(-\chi_2+\chi) \frac{1}{\chi} \right] \nonumber\\
\hspace{-2cm}&\ &\, \times 
\bigg[ \cos(-2\chi_1+\chi+\chi_2) \left( 1+\frac{1}{\chi\chi_1}-\frac{1}{\chi\chi_2}+\frac{1}{\chi_1\chi_2} \right) \nonumber\\ 
\hspace{-2cm}&\ &\, + \sin(-2\chi_1+\chi+\chi_2)\left( -\frac{1}{\chi\chi_1\chi_2}+\frac{1}{\chi_1}-\frac{1}{\chi}-\frac{1}{\chi_2} \right) \bigg].\label{eq:integrand}
\end{eqnarray}
The contribution to the integral 
from the subhorizon $- \chi_1 \gg 1$ is negligible. 
 In the limit of superhorizon $-\chi_1 \ll 1$, we also have
 $-\chi_2 \ll 1 , -\chi \ll 1$. Hence,  the integrand in Eq.(\ref{eq:integrand}) approximately becomes $8/\chi_1 \chi_2 $.
  Thus, the anisotropy can be evaluated as~\cite{Watanabe:2010fh}
\begin{eqnarray}
\frac{\delta \langle in \left| \delta \bar{\phi}_{\bf k}
                             \delta \bar{\phi}_{\bf p} \right| in \rangle}
{\langle 0 \left| \delta \bar{\phi}_{\bf k}
                             \delta \bar{\phi}_{\bf p} \right| 0 \rangle} (\chi) 
&=& \frac{6I}{1-I} \sin ^2 \theta \int^{\chi}_{-1} d\chi_2 \int^{\chi_2}_{-1} d\chi_1 \frac{8}{\chi_1 \chi_2} \nonumber\\
&=& \frac{24I}{1-I} \sin ^2 \theta \ N^2(k),
\end{eqnarray}
where $N(k) \equiv -\ln (-k\eta)$ is the $e$-folding number from the horizon exit.

For the anisotropy in both polarizations of gravitational waves, 
the similar calculations give~\cite{Watanabe:2010fh}
\begin{equation}
\frac{\delta \langle in \left|  \bar{\Gamma}_{\bf k} 
                                \bar{\Gamma}_{\bf p} \right| in \rangle}
{\langle 0 \left|  \bar{\Gamma}_{\bf k} 
                                \bar{\Gamma}_{\bf p} \right| 0 \rangle}
 = \frac{\delta \langle in \left| \bar{G}_{\bf k} \bar{G}_{\bf p} \right| in \rangle}
 {\langle 0 \left| \bar{G}_{\bf k} \bar{G}_{\bf p} \right| 0 \rangle} 
 = 6 I \epsilon_H \sin ^2 \theta \ N^2(k)  \ ,
\end{equation}
where we used the interaction term in the action (\ref{eq:lagbegin})
for $\bar{\Gamma}$ and that in (\ref{J:G}) for $\bar{G}$. 
It is interesting to calculate the cross correlation. 
The leading contribution comes from $H_I^{JG}$ and $H_I^{\phi J}$.
The result is as follows~\cite{Watanabe:2010fh}:
\begin{eqnarray}
\frac{\langle in \big| \delta\bar{\phi}_{\bf k} \bar{G}_{\bf p} \big| in \rangle}
{  \langle 0\big| \delta\bar{\phi}_{\bf k} \delta\bar{\phi}_{\bf p} \big| 0 \rangle}
\simeq - 24 I \sqrt{\frac{\epsilon_H}{1-I}}  N^2(k)  \ .
\label{cross}
\end{eqnarray}
As we will soon see, this might give a detectable number.

\section{How To Test Anisotrpic Inflation}
\label{sec:５}

Now, we are in a position to discuss cosmological implication
of an anisotropic inflationary scenario. As we have listed up in the introduction,
there are many interesting phenomenology in anisotropic inflation. 
Here, we recapitulate the results.
Remember that the anisotropy in the power spectrum is parameterized by
\begin{eqnarray}
  P({\bf k} ) = P(k) \left[ 1 + g_* \sin^2 \theta \right] \ .
\end{eqnarray}
Then, we can predict the following:
\begin{itemize}
\item There exists statistical anisotropy in curvature perturbations
of the order of 
\begin{eqnarray}
 g_s = 24 I N^2(k)  \ .
\end{eqnarray}
\item There exists statistical anisotropy in gravitational waves
of the order of 
\begin{eqnarray}
g_t = 6I \epsilon_H N^2(k)  \ .
\end{eqnarray}
\item These exists the cross correlation 
between scalar perturbations and gravitational waves
of the order of 
$-24 I \sqrt{\epsilon_H} N^2(k) $.
Using the definition of curvature perturbations 
$\zeta = \delta\bar{\phi}/\sqrt{2 \epsilon_H}$, one can translate
the cross correlation (\ref{cross}) between the scalar perturbations and
gravitational waves to that between the curvature perturbations and
gravitational waves normalized by the power spectrum of curvature perturbations:
\begin{eqnarray}
r_c = \frac{\langle in \big| \zeta_{\bf k} \bar{G}_{\bf p} \big| in \rangle}
{  \langle 0\big| \zeta_{\bf k} \zeta_{\bf p} \big| 0 \rangle}
= - 24 \sqrt{2} I N^2(k) \epsilon_H  \ .
\end{eqnarray}
\end{itemize} 
Here, we should note that $I\ll 1$.
Due to the interaction on superhorizon scales, there is an
enhancement factor $N^2(k)$ in the above quantities. 
Because of this enhancement, even when the anisotropy of the spacetime
is quite small, say $\Sigma/H \sim 10^{-7}$ in our example, 
the statistical anisotropy imprinted in primordial fluctuations
can not be negligible in  precision cosmology.   

It is useful to notice that
there exist consistency relations between observables
\begin{eqnarray}
   4 g_t = \epsilon_H \  g_s   \ , \quad r_c = -\sqrt{2} \epsilon_{H} g_s  \ .
\end{eqnarray}
The consistency relations allows us to test anisotropic inflation in a model independent way.
Let us explain how to use it.
It is known that the current observational limit of the statistical anisotropy 
for the curvature perturbations is given by 
$g_s < 0.3$~\cite{Pullen:2007tu}. Now, suppose that we detected $g_s =0.3$.
Then, the consistency relations would give us predictions.
Namely, anisotropic inflation implies the anisotropy in the gravitational waves
\begin{eqnarray}
g_t \simeq 10^{-3}  
\end{eqnarray}
and the cross correlation
\begin{eqnarray}
r_c = -  \sqrt{2} g_s \epsilon_H \sim -  4 \times 10^{-3} \ ,
\end{eqnarray}
where we used $g_{s} \sim 0.3 $ and $\epsilon_H \sim 10^{-2}$.
If these predictions are confirmed by the CMB observations, that must be a strong evidence of anisotropic inflation.  

Apart from the above uses, we can use the observational upper bound  $g_s < 0.3$ to
give a cosmological constraint on the gauge kinetic function. Indeed, we have a constraint 
$24 I N^2 (k) <0.3$. Since $I$ is derived from the gauge kinetic function and the $e$-folding number $N(k)$ can be  
determined once reheating process is clarified, the constraint on $g_s$ implies
the constraint on the gauge kinetic function.

In \cite{Gumrukcuoglu:2010yc}, it is pointed out that the sign of
$g_s$ predicted by our models is different from the observed one. However, it is possible to
modify the model so that the sign of $g_s$ is flipped. 
For example, we can consider two vector fields.
Remarkably, the dynamics of vector fields tends to minimize the anisotropy in
 the expansion of the universe and leading to the orthogonal dyad~\cite{Yamamoto:2012tq}.
 Then, the orthogonal direction to the plane determined by two vectors 
 becomes a preferred direction. In this case, we can expect the sign of
 $g_{s}$ becomes opposite. We can also utilize anti-symmetric tensor fields
to achieve the same aim. 
\begin{figure}
\includegraphics[width=12cm]{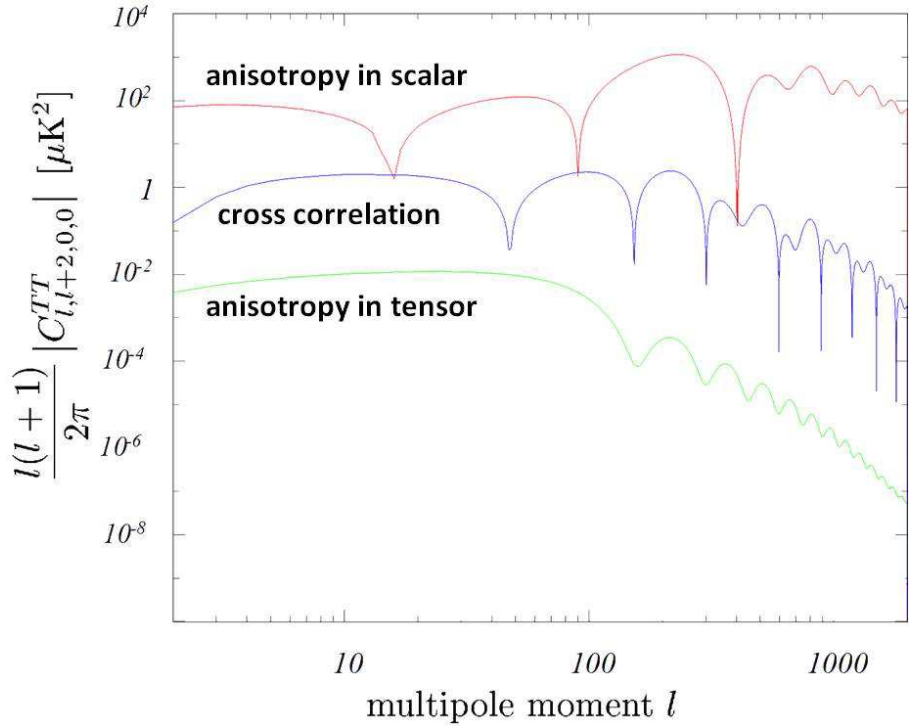}
 \caption{The TT spectra $C^{TT}_{\ell , \ell +2}$ induced by anisotropy in scalar perturbations, that in tensor perturbations and cross correlation. The parameters are chosen as $g_s =0.3,\ r=0.3$.}
 \label{fg:temp}
\end{figure}

We showed in the above how the consistency relations are used to predict the observables. 
So, the next question is how can we see these features in the CMB. 
The answer is that 
the anisotropy related to tensor perturbations induces off-diagonal $TB,EB$ spectra $C_{\ell , \ell +1}$ as well as
 on- and off-diagonal $TT,EE,BB,TE$ spectra $C_{\ell , \ell }$, $C_{\ell , \ell +2}$. Here, we have defined the angular power spectrum
$C_{\ell, \ell'} = <a_{\ell,m} a_{\ell' , m'} >$ with coefficients $a_{\ell m}$ of the spherical harmonic expansion.

\begin{figure}
\includegraphics[width=12cm]{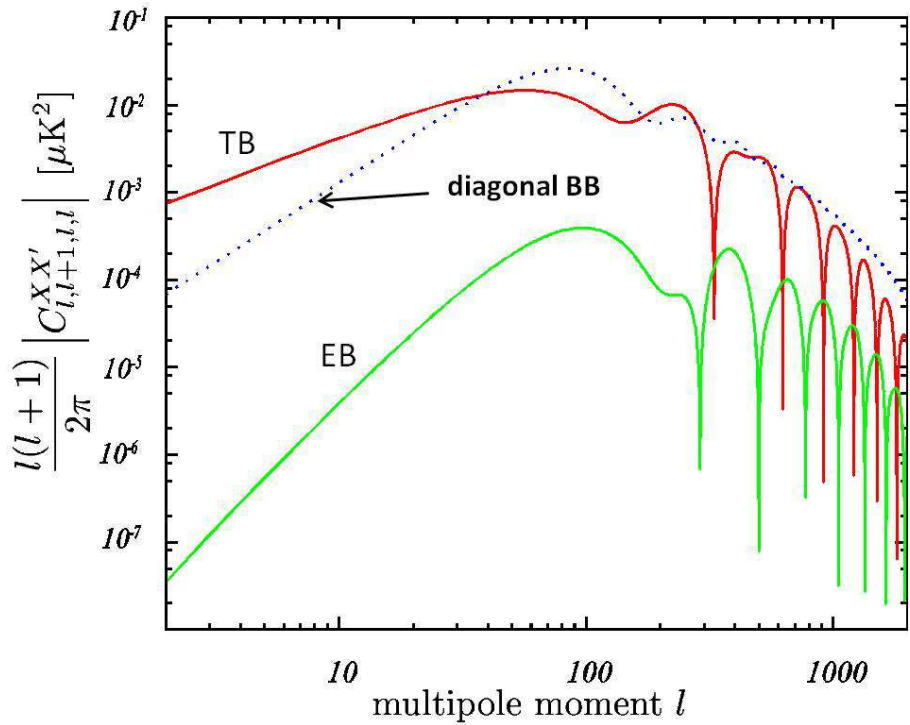}
 \caption{The TB and EB spectra $C^{TB}_{\ell , \ell +1}$, $C^{EB}_{\ell , \ell +1}$ induced by the cross correlation. As a reference, the conventional diagonal BB spectrum induced by isotropic part of the tensor perturbations is plotted with a dotted line. The parameters are taken as $g_s =0.3,\ r=0.3$.}
 \label{fg:tbeb}
\end{figure}

First, we compare the amplitudes of signals induced by the three kind of anisotropy~\cite{Watanabe:2010bu}.
In Fig \ref{fg:temp}, we have depicted off-diagonal $TT$ correlations $C^{TT}_{\ell , \ell +2}$ induced by the anisotropy. 
As for the parameter of anisotropy in scalar perturbations, we adopted the value $g_s =0.3$
as a reference, which is just of the order of a systematic error in WMAP data.
 Note that, according to \cite{Pullen:2007tu}, a signal as small as 2\% can be detected with the PLANCK. 
We also assumed the tensor-to-scalar ratio to be $r = 0.3$. Then the other quantities can be determined by the consistency relations in our model: $r=16\epsilon_H ,\ 4 g_t =\epsilon_H g_s \ , \ r_c= - \sqrt{2 } \epsilon_H g_s $. 
We see that the contributions of the anisotropy in tensor perturbations and the cross correlation are suppressed in comparison to 
that of the anisotropy in scalar perturbations. And, the cross correlation has the contribution next to that of the anisotropy in scalar perturbations. 
This reflects the hierarchy among $g_s$, $r g_t ={\cal O}(g_s \epsilon_H^2), \sqrt{r} r_c = {\cal O}(g_s \epsilon_H )$.
 It is also true for $EE$ and $TE$ spectra.
The ratio between these effects are given by the slow-roll parameter $\epsilon_H$ (or the tensor-to-scalar ratio $r$) 
and does not depend on the value of $g_s$.

Next we consider peculiar signals of anisotropic inflation~\cite{Watanabe:2010bu}.
In Fig \ref{fg:tbeb}, we have depicted examples of $TB$ and $EB$ correlations $C^{TB}_{\ell , \ell +1}$, $C^{EB}_{\ell , \ell +1}$.
 The parameters are again $r=0.3$ and $g_s =0.3$. 
As a reference, the conventional $BB$ diagonal spectrum induced by the isotropic part of tensor perturbations is also plotted with a dotted line.
Note that unlike parity violating cases for which odd parity correlations $C_{ll}^{TB}, C_{ll}^{EB}$ exist~\cite{Lue:1998mq,Alexander:2004wk,Satoh:2007gn,Saito:2007kt},
 our model predicts even parity correlations such as $C_{l,l+1}^{TB}$ as the result of parity symmetry of the system. 
The ratio of $TB$ correlation induced by cross correlation to the isotropic $BB$ correlation is not dependent on $\epsilon_H$ (or $r$) for a fixed value of $g_s$ in our anisotropic inflation model. For the optimistic value of $g_s \sim 0.3$,
 both amplitudes become comparable. This simple order estimation implies 
 that the $TB$ signal could be comparable to that 
of $B$ mode correlation induced by primordial gravitational waves.
Hence, anisotropic inflation can be a potential source of off-diagonal $TB$ correlation, in addition to other effects such as gravitational lensing and our peculiar velocity. Since the current constraints on the $TB/TE$ ratio is of the order 
of $10^{-2}$~\cite{Komatsu:2010fb}, we need to improve
the accuracy by one more order,
which might be achieved by the PLANCK.

\section{Conclusion}
\label{sec:6}

In this review article, we tried to explain how anisotropic inflation 
naturally appears in supergravity models and how to test anisotropic inflationary models
by observations. There are several predictions specific to anisotropic inflation.
There exists the statistical anisotropy in primordial curvature and tensor perturbations.
Furthermore, there exists a cross correlation between the curvature and tensor perturbations
which can be regarded as a kind of the statistical anisotropy.
Most importantly,  there are consistency relations between 
observables  in anisotropic inflationary models. This finding gives rise to a model independent test
of anisotropic inflation. On the other hand, each observable gives the information of the specific model.
Actually, we have already given the first cosmological constraint on the the gauge kinetic function.

The existence of anisotropic inflation can be regarded as a counter example to the cosmic no-hair conjecture.
Indeed, in the presence of the gauge kinetic function, there could be vector-hair
which leads to the anisotropy in the cosmic expansion during inflation. 
Hence, the cosmic no-hair conjecture should be modified appropriately. 
Recently, we have examined inflation with multi-vector fields and found
that inflation tends to minimize their hair~\cite{Yamamoto:2012tq}. 

There are several directions to be explored. It is possible to extend the Bianchi type I model
to other Bianchi type models. 
Theoretically, it is also interesting to embed anisotropic inflation into string theory. 
Observationally, we need to check the data in the CMB more seriously. 
It is worth investigating how to quantify the statistical anisotropy~\cite{Hajian:2003qq,ArmendarizPicon:2005jh,Groeneboom:2008fz,
Akofor:2008gv,ArmendarizPicon:2008yr,arXiv:1107.4304,arXiv:1107.0682,Ma:2011ii}.
In fact, our main message is that the statistical anisotropy is logically allowed,
 natural theoretical models exist, and hence observational check of the statistical anisotropy
needs to be performed. Indeed, there is no apparent reason to respect the non-Gaussianity 
than the statistical anisotropy. 

\ack
I would like to thank Sugumi Kanno, Masashi Kimura, Keiju Murata, Masa-aki Watanabe, Kei Yamamoto, Shuichiro Yokoyama
for collaboration on anisotropic inflation. 
I am grateful to  Konstantinos Dimopoulos, Hassan Firouzjahi, Sigbjorn Hervik, M.M.Sheikh-Jabbari, Hideo Kodama, David Lyth, 
Kei-ichi Maeda, Azadeh Maleknejad, Shinji Mukohyama, Marco Peloso, Valery Rubakov, Misao Sasaki
for fruitful discussions. 
This work was supported in part by the
Grant-in-Aid for  Scientific Research Fund of the Ministry of 
Education, Science and Culture of Japan No.22540274, the Grant-in-Aid
for Scientific Research (A) (No.21244033, No.22244030), the
Grant-in-Aid for  Scientific Research on Innovative Area No.21111006,
JSPS under the Japan-Russia Research Cooperative Program,
the Grant-in-Aid for the Global COE Program 
``The Next Generation of Physics, Spun from Universality and Emergence".

\end{document}